\documentclass[final,5p]{elsarticle}
\usepackage{graphicx}\usepackage{multirow}\usepackage{amsmath,amssymb,amsfonts}\usepackage{amsthm}\usepackage{mathrsfs}\usepackage[title]{appendix}\usepackage{xcolor}\usepackage{booktabs}\usepackage{longtable}\usepackage{array}\usepackage{tabularx}\usepackage{listings}
\usepackage[T1]{fontenc}\usepackage{lmodern}\usepackage{microtype}\usepackage{xurl}\usepackage{hyperref}\usepackage{newunicodechar}
\newunicodechar{±}{\ensuremath{\pm}}\newunicodechar{²}{\textsuperscript{2}}\newunicodechar{µ}{\textmu}\newunicodechar{·}{\ensuremath{\cdot}}\newunicodechar{×}{\ensuremath{\times}}\newunicodechar{Δ}{\ensuremath{\Delta}}\newunicodechar{α}{\ensuremath{\alpha}}\newunicodechar{β}{\ensuremath{\beta}}\newunicodechar{π}{\ensuremath{\pi}}\newunicodechar{ρ}{\ensuremath{\rho}}\newunicodechar{τ}{\ensuremath{\tau}}\newunicodechar{φ}{\ensuremath{\varphi}}\newunicodechar{ᵃ}{\textsuperscript{a}}\newunicodechar{ᵇ}{\textsuperscript{b}}\newunicodechar{ᶜ}{\textsuperscript{c}}\newunicodechar{–}{\textendash}\newunicodechar{—}{\textemdash}\newunicodechar{…}{\ldots}\newunicodechar{⁴}{\textsuperscript{4}}\newunicodechar{⁵}{\textsuperscript{5}}\newunicodechar{⁶}{\textsuperscript{6}}\newunicodechar{⁷}{\textsuperscript{7}}\newunicodechar{⁸}{\textsuperscript{8}}\newunicodechar{⁻}{\textsuperscript{-}}\newunicodechar{ℝ}{\ensuremath{\mathbb{R}}}\newunicodechar{→}{\ensuremath{\rightarrow}}\newunicodechar{↔}{\ensuremath{\leftrightarrow}}\newunicodechar{∅}{\ensuremath{\emptyset}}\newunicodechar{∈}{\ensuremath{\in}}\newunicodechar{−}{\ensuremath{-}}\newunicodechar{∪}{\ensuremath{\cup}}\newunicodechar{≈}{\ensuremath{\approx}}\newunicodechar{≤}{\ensuremath{\leq}}\newunicodechar{≥}{\ensuremath{\geq}}\newunicodechar{á}{\'a}\newunicodechar{ä}{\"a}\newunicodechar{é}{\'e}\newunicodechar{ó}{\'o}\newunicodechar{ü}{\"u}\newunicodechar{Ł}{\L}\newunicodechar{ś}{\'s}
\providecommand{\tightlist}{\setlength{\itemsep}{0pt}\setlength{\parskip}{0pt}}\providecommand{\passthrough}[1]{#1}
\hypersetup{pdftitle={OntoKG-EQ: Evidence-boundary auditing for analytical knowledge graphs},pdfauthor={Furqan Nasir, Muhammad Atif Saeed, Muhammad Ehsan, Sher Jeel Ahmad, Abdul Moiz Altaf},pdfkeywords={Knowledge graphs, Ontology engineering, Competency questions, Provenance, SHACL, Explainable querying}}
\makeatletter\def\ps@pprintTitle{\let\@oddhead\@empty\let\@evenhead\@empty\def\@oddfoot{\footnotesize\itshape arXiv preprint, version 2\hfill November 2026}\let\@evenfoot\@oddfoot}\makeatother

\begin{document}
\begin{frontmatter}
\title{OntoKG-EQ: Evidence-boundary auditing for analytical knowledge graphs}

\author[cusit,fast]{Furqan Nasir\corref{cor1}}
\ead{furqan.nasir@cusit.edu.pk}
\author[fast]{Muhammad Atif Saeed}
\ead{atif.saeed@isb.nu.edu.pk}
\author[cusit]{Muhammad Ehsan}
\ead{me@cusit.edu.pk}
\author[uet]{Sher Jeel Ahmad}
\ead{Sherjeel4all@gmail.com}
\author[cusit]{Abdul Moiz Altaf}
\ead{moizaltaf8080@gmail.com}
\affiliation[cusit]{organization={City University of Science and Information Technology (CUSIT)},city={Peshawar},country={Pakistan}}
\affiliation[fast]{organization={National University of Computer and Emerging Sciences (FAST-NUCES)},city={Islamabad},country={Pakistan}}
\affiliation[uet]{organization={University of Engineering and Technology (UET)},city={Peshawar},country={Pakistan}}
\cortext[cor1]{Corresponding author.}

\begin{abstract}
An analytical knowledge-graph result can be supported in its ambient graph while its delivered explanation omits deciding observations. OntoKG-EQ connects competency-question requirements to an executable audit of ambient observations, declared finding support and delivered evidence. A bounded declarative interpreter distinguishes positive, false and unknown numerical status, with same-witness unit readiness reported separately. A public-data application reconstructs England's August 2026 provider-level referral-to-treatment statistical screen. Independent source recomputation reconciles 532 providers with the publisher workbook: 91 meet the inclusive 92\% threshold and 441 do not. One additional roster entry remains unavailable rather than zero-filled. All 533 application frames agree with an independently implemented standards-based validation and availability baseline. Transfer requires a disclosed source adapter and arithmetic compilation, but no change to the core interpreter. Twenty-six constructed analytical cases yield 78 status/readiness agreements. Actual financial release replay reproduces two omitted-volume defects: support and delivery suffice for 19 of 21 original findings and all 21 repaired findings after interface adaptation. A separate replay agrees on 487 overlapping financial records. Source audits expose scaling errors and unresolved comparator and temporal assumptions, so financial examples do not establish authenticated market effects. The contribution is a reproducible analytical audit interface and source-reconciled application, not new sufficiency logic or first evidence-boundary isolation. Historical ratings measure preference only; objective human audit benefits and independent production deployment remain untested.

\end{abstract}


\end{frontmatter}

\hypertarget{introduction}{%
\section{Introduction}\label{introduction}}

Analytical knowledge graphs combine heterogeneous observations with executable queries, but reproducing an answer requires more than recovering its result entities. An auditor must identify the query conditions, the observations that decided those conditions, their temporal context, and their sources. OntoKG-EQ investigates the following question: \emph{How can competency-question requirements govern the construction and validation of analytical knowledge graphs so that query results carry inspectable supporting evidence?}

\begin{figure*}[t]
\centering
\includegraphics[width=0.85\textwidth]{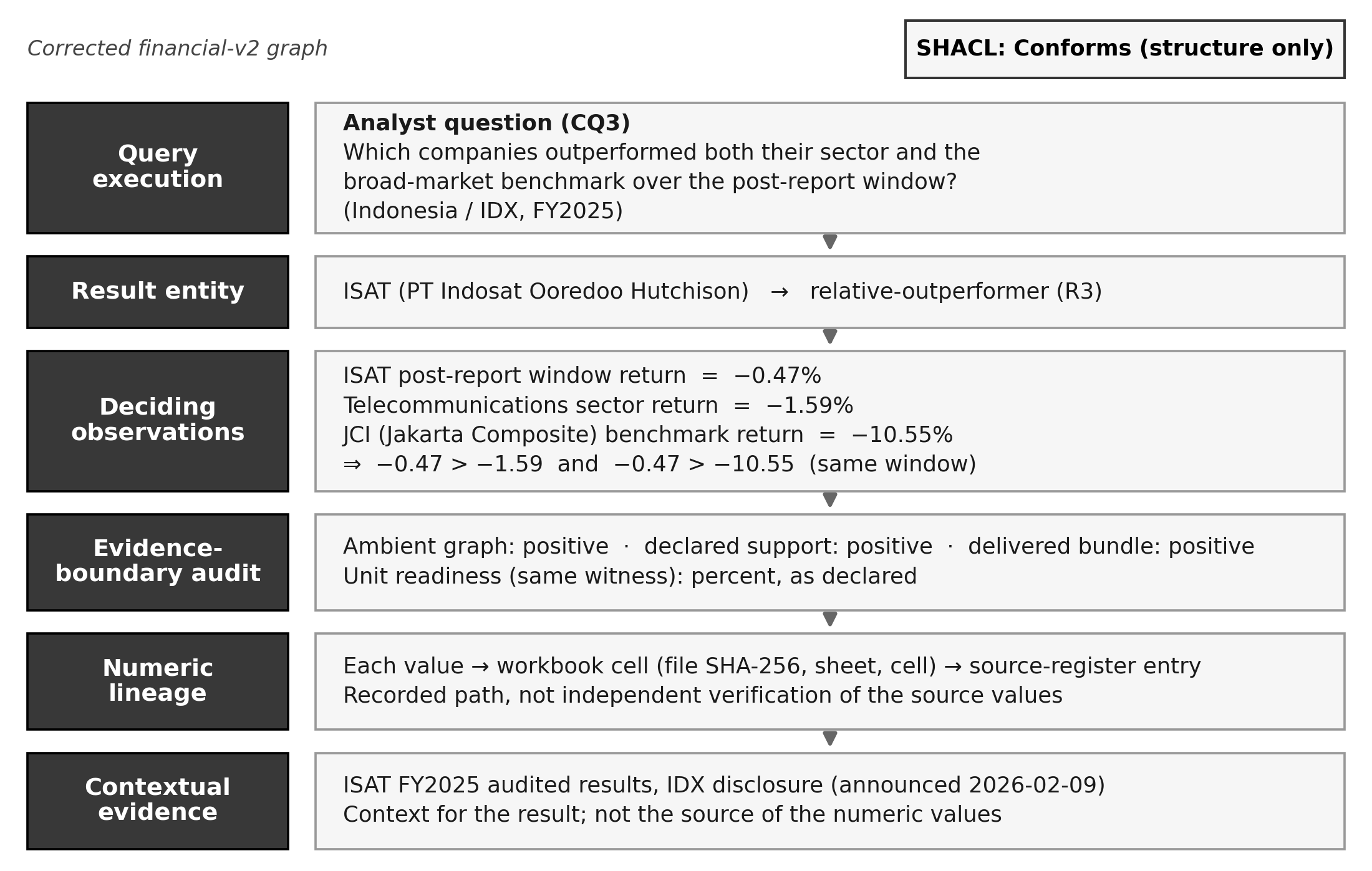}
\caption{A worked example (Indonesia, CQ3, corrected financial-v2 graph). ISAT is returned as a relative outperformer because its window return exceeds both same-window comparators. The evidence-boundary audit finds the condition positive in the ambient graph, the declared support and the delivered bundle, with declared units on the same witness. Numeric values trace to recorded workbook cells; the FY2025 IDX disclosure is contextual evidence and does not establish those values. SHACL conformance concerns structure only. Values are demonstrator fixtures, not authenticated market returns.}\label{fig:worked-example}
\end{figure*}

Equity analysis provides the original case study. Analysts compare reported fundamentals with market response, examine exchange-rate associations, compare company returns with sector and benchmark returns, and inspect announcement-window activity. These tasks combine numerical computation with requirements for temporal alignment and source traceability. Figure~\ref{fig:worked-example} illustrates the intended evidence path for one relative-outperformance result. The financial setting supplies concrete requirements and evaluation data; the reusable design concerns the organization of requirements, semantic artifacts, and result evidence.

The additional public-data application asks an externally specified question
about incomplete referral-to-treatment pathways in England. Unlike the aligned
financial fixtures, it starts from a publisher's raw aggregate reporting schema
and independently reconciles the deciding counts with the publisher's workbook.
This tests whether the audit interface can retain explicit source, entity,
reporting-period and evidence-boundary distinctions in a nonfinancial task.
It does not assume that technical portability establishes practitioner uptake.

We treat this as a problem of \emph{data and knowledge
engineering}, not prediction. The goal is not to forecast prices or
recommend trades. It is to make analyst-oriented queries over
heterogeneous market data \textbf{reproducible, evidence-linked,
temporally explicit, structurally valid, and inspectable}. Every
materialized finding can be inspected together with its explicit supporting observations and temporal context. Source attribution is checked separately; a company disclosure linked as context does not establish the source of every numerical observation.

Our approach builds on established knowledge-graph foundations and
surveys \cite{ref1,ref2}, SPARQL query semantics \cite{ref3}, and the
explainable-AI literature \cite{ref4}, adapting them to the bounded,
provenance-first setting of analyst querying. The novelty question concerns the explicit connection between CQ requirements, executable analytical conditions, and inspectable result evidence; the individual standards and the general idea of provenance-based explanation are established.
Financial-domain ontologies such as FIBO model the concepts of finance
comprehensively, but do not, on their own, provide a bounded, executable
path from a parameterised query to the evidence and provenance that
justify each result. Recent financial knowledge graphs emphasise
large-scale, often LLM-driven, extraction, and explicitly flag
interpretability and provenance as open challenges \cite{ref5,ref6}.
Graph-based retrieval-augmented generation (GraphRAG) grounds
language-model answers in retrieved subgraphs, but its faithfulness is
estimated rather than guaranteed, because a generated answer may not be
entailed by the retrieved evidence.

We present \textbf{OntoKG-EQ}, a knowledge-based system that
operationalises explainable analyst querying over equity data. OntoKG-EQ
is built around five frozen competency questions (CQ1--CQ5) that bound
its scope: no ontology class, property, SHACL shape, derived metric, or
inference rule exists unless it is required to answer one of them. Given
a market's data, the system (i) materialises it into a provenance-aware
RDF knowledge graph aligned to a compact core ontology; (ii) computes
the derived analytical metrics the questions require (window returns,
year-on-year growth, sector and benchmark returns, exchange-rate
association, and event-window abnormal return and volume); (iii)
validates structure with SHACL; (iv) answers each competency question
with SPARQL that computes its stated condition; (v) applies inference
rules that derive typed, queryable analytical findings (e.g.,
relative-outperformer, fundamentals--market divergence, abnormal-event
reaction); and (vi) generates, automatically and for every result, an
evidence bundle that exposes the inspectable path from the executed
query to its supporting observations, evidence items, official sources,
and provenance. Deterministic rendering avoids stochastic changes during transcription, but consistency and completeness depend on the correctness of the selected observations and links.

A secondary claim concerns \textbf{reuse of the apparatus} over
structurally aligned datasets, not arbitrarily heterogeneous reporting
structures. We demonstrate it on
three market-specific demonstrator datasets: the Pakistan
Stock Exchange (PSX, KSE-100 index), Bursa Malaysia (MSX, FBM KLCI
index), and the Indonesia Stock Exchange (IDX, Jakarta Composite). The
core ontology, SHACL shapes, the five competency-question queries, and
the inference rules are reused byte-identical across all three markets.
Only the data and a small per-market adapter configuration change. Every market is SHACL-conformant under the declared shapes. The revision separately checks finding dependencies and workbook-level lineage.

The contributions are threefold:
\begin{enumerate}
\item An evidence-boundary audit protocol that evaluates an analytical condition independently over ambient observations, declared finding support and the actually delivered explanation bundle, without borrowing omitted observations across these boundaries.
\item An RDF/SPARQL implementation with preserved financial replay and a bounded, declaratively configured audit interpreter. Witness sufficiency, metadata readiness, recorded source paths and external source truth remain distinct properties; the interpreter is not a general provenance calculus.
\item A source-reconciled public NHS statistical-screen application with explicit eligibility and missingness, alongside a diagnostic benchmark with matched SHACL-SPARQL comparison and paired financial replay. The NHS application uses a new adapter and arithmetic compilation, rather than claiming unchanged whole-pipeline reuse. Existing complete-query, withholding and source audits provide complementary evidence rather than additional independent samples. Historical participant ratings provide preference evidence only.
\end{enumerate}
The method instantiates established semantic technologies. Its value must be assessed through the explicit organization and verifiability of requirements and evidence, rather than by claiming that provenance or deterministic explanations are unique to knowledge graphs. The original and revision code and data are publicly released (see Data and code availability).

\hypertarget{related-work}{%
\section{Related work}\label{related-work}}

OntoKG-EQ sits at the intersection of four lines of work. We review each
line of work and state precisely what OntoKG-EQ adds.

\hypertarget{financial-ontologies-and-knowledge-graphs}{%
\subsection{Financial ontologies and knowledge
graphs}\label{financial-ontologies-and-knowledge-graphs}}

The Financial Industry Business Ontology (FIBO) \cite{ref7,ref8} is the
de-facto OWL standard for finance, governed by the EDM Council and OMG
and paired with triplestores for inference and property-path querying.
FIBO targets broad \emph{conceptual coverage}. Recent work extends it in
specific directions: FinCaKG-Onto, for instance, uses FIBO as a scaffold
to depict financial expertise as a causality knowledge graph \cite{ref9}.
OntoKG-EQ takes the opposite stance. It is a bounded,
competency-question-scoped core in which every term exists to answer one
of five frozen analyst questions, and it provides an \emph{executable,
inspectable analytical-querying capability} rather than a domain
vocabulary.

A fast-growing parallel line builds financial knowledge graphs by
automated extraction. FinReflectKG constructs a graph from SEC filings
with agentic, reflection-driven extraction and a rule-, statistical-,
and LLM-as-judge evaluation pipeline \cite{ref5}. FinKario adds
event-enhanced automated construction coupled to a two-stage graph
retrieval strategy for stock-trend prediction \cite{ref10}. A comprehensive
survey of the field \cite{ref6} catalogues applications such as fraud,
credit risk, anti-money-laundering, and compliance, while explicitly
naming interpretability and provenance as open challenges. Notably, this
concern is not new to the extraction community itself: Kertkeidkachorn
and Ichise argue that automatically constructed financial graphs,
lacking a well-defined ontology, suffer degraded reasoning and quality,
and respond with FinKG, an expert-verified core ontology \cite{ref11}.
Financial KG construction also includes explicit source attribution and
multidimensional evaluation. FinReflectKG--EvalBench separates faithfulness,
precision, relevance and comprehensiveness for financial extraction
\cite{jwsFinEval}. Consequently, evidence and governance are not absent from
this research line. OntoKG-EQ targets a different operational task: checking
the typed observations and temporal bindings that suffice for a particular
analytical condition. Competency questions have also been used to evaluate
financial graphs \cite{ref12}; our frozen question set bounds the demonstrator
and its artifacts without establishing a new general ontology-engineering
principle.

\hypertarget{competency-question-driven-ontology-engineering}{%
\subsection{Competency-question-driven ontology
engineering}\label{competency-question-driven-ontology-engineering}}

Competency questions (CQs) are a long-standing device for scoping and
validating ontologies \cite{ref13,ref14}, embedded in ontology-engineering
methodologies such as NeOn \cite{ref15}. A recent survey of 63 practitioners
confirms that CQs are used mainly to define scope and evaluate a
conceptualization, but also that engineers still lack guidance for
writing and managing them \cite{ref16}, and Keet and Khan argue that CQs
serve manifold roles across the engineering lifecycle beyond mere
fact-seeking \cite{ref13}. In practice, however, CQs are seldom published
alongside the ontologies they shaped \cite{ref17}, which leaves their
governing intent implicit. There is also active work on formalising CQs
as SPARQL/SPARQL-OWL queries \cite{ref18} and tooling that mints CQ→SPARQL
pairs for question answering over knowledge graphs \cite{ref19}.
Recent work makes the requirement-to-ontology connection explicit:
CQ4OE publishes CQ-to-term and CQ-to-axiom provenance and distinguishes
partial from full requirement coverage \cite{jwsCQ4OE}; MASEO records CQ
and validation rationale during ontology construction \cite{jwsMASEO}.
OntoKG-EQ therefore does not introduce CQ-based provenance. It evaluates
a bounded runtime contract connecting analytical metric roles, selection
conditions, matching windows and sufficient finding evidence. Documented
CQ labels and executable sufficiency checks establish different properties.
The five frozen questions remain a reproducibility boundary, not proof
that the domain requirements are exhaustive or unambiguous.

\hypertarget{provenance-and-structural-validation}{%
\subsection{Provenance and structural
validation}\label{provenance-and-structural-validation}}

Semantic answer explanation is an established research topic. Inference Web represents justifications for retrieved and inferred answers through provenance and derivation information \cite{jwsInference}. Dividino et al.\ formalize and implement querying RDF together with source, trust, uncertainty, and other meta-knowledge \cite{jwsMeta}. Consequently, per-answer explanation and queryable provenance are not new capabilities introduced by OntoKG-EQ.

The PROV model provides interoperable provenance representation \cite{ref20,jwsProvRationale}; it does not establish that source records or reported values are true. SHACL provides declared graph constraints and validation reports \cite{ref21}. Recent work studies validation efficiency and semantics \cite{ref22,ref23,ref24}, and provenance of graph construction \cite{ref25}. Native Provenance Computation for SPARQL provides fine-grained provenance of query results, including non-monotonic queries \cite{jwsNPCS}. Analytical conditions can be encoded in queries or SHACL-SPARQL; they are not inherently beyond these mechanisms.

The closest evidence-oriented predecessors impose a stronger novelty boundary.
Delva et al.\ establish sufficient SHACL neighborhoods and shape fragments,
including nonconformance explanations \cite{jwsShapeProvenance}.
Fierro's Shifty returns satisfaction traces and failure witnesses, including
Boolean alternatives and an application-input example with unit requirements
\cite{jwsShifty}. Ontological Knowledge Blocks combine declarative obligations,
evidence and executable constraints \cite{jwsOKB}; manufacturing trust
infrastructure combines temporal state, provenance and decision records
\cite{jwsManufacturing}. Thus neither sufficient evidence, branch-sensitive
justification, declarative contracts nor temporal-provenance integration alone
is our novelty claim. We operationalize a narrower audit task: independently
check the ambient observations, declared finding support and actually delivered
bundle under the same analytical condition, while retaining separate metadata
and source qualifications. Whether this organization provides an engineering
advantage must be evaluated, not inferred from standards choice.

Scope separation itself also has direct predecessors. SHACL-DS evaluates
declared graph views; its railway study shows that merging graphs can mask a
missing reference by borrowing it from another contributor
\cite{jwsSHACLDS}. ESGlass separates internal evidence assembly from public
rendering and proposes incomplete-bundle diagnostics \cite{jwsESGlass}; it is
an unreviewed concept paper rather than an executed equivalent comparator in
the inspected source. These works prevent a first-boundary-isolation claim.
Our distinction is the concrete application-specialized analytical interface
and its executed paired tests, not discovery of a previously unknown need for
scoped evidence or abstention.

\hypertarget{explainable-querying-graphrag-and-faithfulness}{%
\subsection{Explainable querying, GraphRAG and
faithfulness}\label{explainable-querying-graphrag-and-faithfulness}}

Graph-based retrieval-augmented generation \cite{ref26}, which extends
retrieval-augmented generation \cite{ref27}, retrieves subgraphs to ground
language-model answers, and a rich family of systems now refines this
idea: HybridRAG combines graph and vector retrieval for financial
question answering over earnings calls \cite{ref28}, and Think-on-Graph 2.0
interleaves graph and text retrieval to deepen reasoning \cite{ref29}. There
is also explicit work on explaining KG-RAG \cite{ref30} and on detecting
hallucinations in retrieval-augmented generation \cite{ref31}. The recurring
difficulty across all of these is \emph{faithfulness}: even with
retrieval, a generated answer may not be entailed by the retrieved
context. Recent evaluation work makes the point sharply. Ahmad and Khan
show that lexical metrics such as BLEU and ROUGE correlate only weakly
with structural grounding, and that removing graph retrieval preserves
lexical accuracy while eliminating traceable evidence altogether
\cite{ref32}. Faithfulness, in other words, is a structural property that
text-overlap scores do not capture. This is also increasingly recognised
outside the evaluation literature: Sequeda et al., writing from
enterprise practice, argue that knowledge graphs provide the formal
frame to check a generated query's validity, the foundation for
explaining a result, and access to governed, trusted data, precisely the
roles a bare language model cannot fill \cite{ref33}.

Recent ontology-guided GraphRAG evaluates canonicalisation, traceability and verifiability separately \cite{jwsNeuroGraph}. This prevents a blanket contrast in which generative graph systems lack traceable sources. OntoKG-EQ uses deterministic query execution and rendering, conditional on correct queries, evidence selection and rendering. It does not resolve source truth by construction. Work on non-generative answering \cite{ref34} is a relevant predecessor. The historical transcription panel holds retrieval fixed and is excluded from the revised main quantitative evaluation because generated answer texts were not archived.

\hypertarget{positioning-and-novelty}{%
\subsection{Positioning and novelty}\label{positioning-and-novelty}}

Table~\ref{tab:t1} identifies specific predecessors and the relationship of their verified objectives to this system. An unreported capability is not evidence that a system lacks it. The scope of the present comparison is the analytical workflow implemented here, rather than firstness of explanation or provenance.

\begin{table}[htbp]
\caption{Named predecessors and the contribution boundary. NR means not established by the inspected source, not absent.}\label{tab:t1}
\centering\footnotesize
\begin{tabularx}{\linewidth}{XXX}
\toprule
Work & Established mechanism & Relationship to OntoKG-EQ \\
\midrule
Inference Web \cite{jwsInference} & Provenance and derivations explaining answers & Direct predecessor for individual-answer justification; current CQ-to-metric contract NR \\
Dividino et al.\ \cite{jwsMeta} & Representation and querying of RDF meta-knowledge & Direct predecessor for queryable source/provenance information; current analytical contract NR \\
PROV rationale \cite{jwsProvRationale} & General provenance representation and design & Foundation for attribution, not a competing equity-analysis system \\
Wisniewski et al.\ \cite{jwsCQ} & Published CQs and SPARQL-OWL formalizations & Prior executable requirement artifacts \\
CQ4OE / MASEO \cite{jwsCQ4OE,jwsMASEO} & CQ-to-ontology provenance and construction rationale & Prior design traceability; distinct from runtime analytical witness checks \\
NPCS \cite{jwsNPCS} & Fine-grained SPARQL result provenance & Generic derivation provenance; no untested claim of inferior domain validation \\
SHACL neighborhoods \cite{jwsShapeProvenance} & Sufficient subgraphs and nonconformance provenance & Direct predecessor; no new sufficiency theorem claimed \\
Shifty \cite{jwsShifty} & Passing/failing evidence, Boolean witnesses, application-input contracts & Direct predecessor; three-pool analytical audit is the evaluated task, not an expressiveness separation \\
OKB / manufacturing trust \cite{jwsOKB,jwsManufacturing} & Executable obligations, evidence and temporal decision provenance & Prior integration mechanisms; no first-contract or first-temporal-provenance claim \\
SHACL-DS \cite{jwsSHACLDS} & Scoped graph views and observed merged-graph masking & Direct predecessor for evidence-scope isolation; our observation pools retain ambient metadata \\
ESGlass \cite{jwsESGlass} & Conceptual internal/public evidence separation and incomplete-bundle tests & Prior agenda; implemented analytical diagnostics, not conceptual firstness, distinguish this study \\
Ontology-guided GraphRAG \cite{jwsNeuroGraph} & Canonicalisation and traceability audits & Relevant source-verification distinction; different end-to-end answering task \\
OntoKG-EQ & CQ-linked analytical contracts and sufficient-witness replay & Bounded role, window and predicate checks; source-cell linkage evaluated separately \\
\bottomrule
\end{tabularx}
\end{table}

Executable competency questions also have established precedents. Wisniewski et al.\ publish and analyze CQs with SPARQL-OWL formalizations \cite{jwsCQ}. Our traceability matrix documents the association of requirements with artifacts; it does not prove ontological minimality or correct interpretation of every natural-language question. We assess the composition through declared constraints, result dependencies, numerical checks, and reproducible artifact reuse.

The resulting research question is not whether provenance can explain an answer,
but whether a delivered analytical explanation contains an admissible witness
without borrowing omitted observations from its ambient graph. A finding can
be numerically true in the graph while its delivered bundle is incomplete or
contains only nonpassing candidates. Conversely, deleting one observation from
a bundle need not invalidate it when an independent sufficient branch remains.
These are diagnostic distinctions to test against a semantically matched
encoding, not evidence that existing validation languages cannot express them.

\hypertarget{preliminaries-and-formal-model}{%
\section{Preliminaries and formal
model}\label{preliminaries-and-formal-model}}

\hypertarget{frozen-competency-questions}{%
\subsection{Frozen competency
questions}\label{frozen-competency-questions}}

The scope of OntoKG-EQ is fixed by five competency-question (CQ)
families, frozen before modelling. They are the functional requirements
the system must answer and the governance boundary for every modelling
decision: no ontology term, data field, SHACL shape, derived metric, or
inference rule is admitted unless at least one CQ requires it.

\begin{itemize}
\tightlist
\item
  \textbf{CQ1. Fundamentals versus market response.} Which companies
  show stronger reported fundamentals but a weaker subsequent market
  response over a defined post-reporting window?
\item
  \textbf{CQ2. Exchange-rate context and market behaviour.} How is
  movement in a selected local currency pair associated with company-,
  sector-, or benchmark-level market measures over a predefined window?
\item
  \textbf{CQ3. Relative outperformance versus sector and benchmark.}
  Which companies outperform their sector comparator and the selected
  broad-market benchmark over the same period?
\item
  \textbf{CQ4. Official announcements versus market reaction.} Which
  official announcements and associated disclosures are coincident,
  within short event windows, with abnormal return or trading-volume
  activity in short event windows such as (−1, +1) and (−3, +3) trading
  days?
\item
  \textbf{CQ5. Explainability and provenance.} What evidence,
  intermediate relations, metric values, temporal context, sources, and
  provenance explain why an entity appears in an executed CQ1--CQ4
  result?
\end{itemize}

CQ1--CQ4 are \emph{analytical} questions over the data. CQ5 is the
\emph{explainability} question that the other four must be answerable
against.

\hypertarget{core-constructs}{%
\subsection{Core constructs}\label{core-constructs}}

Let the core ontology be \texttt{O\ =\ (C,\ P)} where C is the set of
classes and P the object and datatype properties, and let \texttt{S} be
a separate validation schema (the SHACL shape set) layered over
\texttt{O}. Both are \emph{CQ-bounded}: each element of
\texttt{C\ ∪\ P\ ∪\ S} is justified by at least one
\texttt{q\ ∈\ Q\ =\ \{CQ1,\ \ldots{},\ CQ5\}}.

\textbf{Observation.} An observation is a tuple
\texttt{o\ =\ (e,\ m,\ v,\ τ,\ X,\ r)} where e is the observed entity (a
Company, an IndustrySectorClassifier, a MarketIndex, or a currency
pair), m a metric name, \texttt{v\ ∈\ ℝ} its value, τ a temporal context
(a ReportingPeriod, an AnalysisWindow, or an EventWindow), X a possibly
empty set of supporting EvidenceItems, and r a ProvenanceRecord. Three
subclasses specialise the entity and temporal context:
FundamentalObservation, MarketObservation, and ExchangeRateObservation.

\textbf{Derived metrics and inference rules.} A derived metric
\texttt{φ\ :\ (2\^{}Obs,\ W)\ →\ Obs} maps a set of base observations
and a window to a new, first-class (derived) observation, and OntoKG-EQ
defines nine CQ-justified metrics. An inference rule
\texttt{ρ\ :\ Pattern\ →\ Finding} maps a graph pattern encoding a CQ
condition over these metrics to a typed AnalyticalFinding \texttt{f}.
The nine-metric list and the finding tuple
\texttt{f\ =\ (e,\ t,\ ρ\_id,\ q,\ O\_f)} are given in Online Resource 1
(Section D).

\textbf{Evidence bundle and explanation.} For a result entity e, an
evidence bundle \texttt{b\ =\ (e,\ O\_b,\ X\_b)} collects the supporting
observations \texttt{O\_b} and evidence items \texttt{X\_b}. A query
execution \texttt{qe\ =\ (q,\ id,\ π,\ b)} records an executed CQ
instance with parameters \texttt{π} bound to \texttt{b}. The
\textbf{explanation} of \texttt{e} is the closure
\texttt{E(e)\ =\ \{\ (o,\ m,\ v,\ τ,\ x,\ src,\ R)\ :\ o\ ∈\ O\_b,\ x\ ∈\ evidence(o),\ src\ =\ source(x)\ \}},
where \texttt{R\ =\ prov(o)\ ∪\ prov(x)} is the set of provenance
records linking observation o to its supporting evidence item x. R
denotes a set, distinct from the single ProvenanceRecord r of an
individual finding above; restricting x to evidence(o) avoids forming
unrelated observation--evidence pairs.

\hypertarget{quality-properties}{%
\subsection{Quality properties}\label{quality-properties}}

We distinguish five properties. \textbf{Structural conformity} means that a particular graph snapshot satisfies the declared SHACL shapes; this is relative to the coverage of those shapes. \textbf{Evidence availability} means that an answer resolves to an observation and a source. \textbf{Dependency completeness} is stronger: the explanation includes all observations needed to check the actual query or rule condition, including matching temporal and comparator bindings. \textbf{Analytical correctness} requires independent verification of the derived quantities. \textbf{Source fidelity} requires checking that observations and source associations faithfully represent the originating records.

For a fixed graph snapshot, deterministic rendering is consistent with its selected evidence only if the renderer copies and labels those values correctly. A complete explanation additionally assumes correct query/rule implementations and complete dependency selection. Neither determinism nor SHACL conformance proves source truth or analytical correctness. In particular, a well-typed incorrect number or an unrelated but well-formed source may satisfy all declared shapes. The evaluation therefore treats these properties separately. A non-empty evidence path alone does not establish a complete justification.

\subsection{Executable finding-evidence contracts}\label{executable-contracts}
Each finding classification has a versioned contract specifying its metric
roles, entity relations, temporal-context predicates, expected units and
selection condition. A machine-readable manifest identifies the related
queries, builder, rule implementation, documented rules, shapes and ontology.
Artifact hashes identify versions; a hash match does not establish semantic
correctness. Design provenance connects requirements to these artifacts,
while execution provenance connects a finding to observations and source cells.

Let $K_q$ denote a declared contract and $B_f$ the observations included in
the actual evidence bundle for finding $f$, restricted to its declared support.
A sufficient witness $W\subseteq B_f$ assigns compatible observations to the
contract's roles and makes its predicate true. CQ1 requires positive growth
and a company statistic below its same-window benchmark. CQ3 requires the
company statistic to exceed both same-window comparators. The R2 contract
checks the FX-sensitive classification only; it does not validate every field
of the broader CQ2 answer. R4 permits either a passing event-return branch or
a passing volume branch. Both operands need not be necessary evidence for an
OR condition. Expected units are contract assumptions when the graph lacks
explicit unit declarations.

The checker separately replays declared finding support and observations
available in the correctly targeted bundle, without importing the production
SPARQL rules. Missing roles produce an explicit incomplete status; a complete
role assignment that fails the predicate produces a predicate-false status.
A missing bundle dependency can therefore be detected even when the finding's
support links remain intact. The operational complete-bundle rate is the share
of inspected positive findings whose actual bundle contains at least one
sufficient witness. This is relative to the declared predicate, not proof of
global minimality, recall of all correct answers, or complete domain knowledge.

For a source-linked witness, every recorded dependency cell of each witness
observation must resolve to one registered source with a matching identifier.
This checks path integrity rather than original-source truth or completeness
of upstream computation. R1 reporting-period-to-window anchoring and R4
finding-to-event anchoring remain unverified where explicit links are absent.
The checker writes separate witness/status overlays and preserves the archived
graphs. Contract coverage is restricted to these finding classifications;
it is not a validator for arbitrary SPARQL or all five complete CQ outputs.

\paragraph{Operational interface and reuse boundary.}
The checker receives a graph snapshot, a finding identifier, its selected
versioned contract and optionally a separate lineage graph. It enumerates typed
metric-role candidates under declared entity/window relations and evaluates
numeric conditions with Decimal arithmetic. Outputs separate full-graph truth,
finding-support sufficiency, actual-bundle status, witness identifiers and
recorded source-path status; unknown assumptions are not promoted to verified
source facts. The preserved financial checker has a manually specified role
vocabulary and four predicate implementations. An additive interpreter instead
reads bounded role bindings and numerical branches from configuration; the two
implementations are distinguished in the evaluation. Adding a domain still
requires authored mapping, contract and test work. Neither implementation is a
general provenance calculus or an empirically validated cross-domain analytical
system.

\subsection{Evidence-boundary auditing}\label{boundary-method}
For a fixed graph snapshot $G$, entity $e$ and contract $K$, let $A$ be the
ambient observation pool, $S$ the finding-declared support and $B$ the declared
bundle membership. The delivered pool is $D=B\cap S$ only when the bundle
association identifies the finding's entity unambiguously; otherwise $D$ is
empty and an association diagnostic is returned. Extraneous bundle members
are reported rather than admitted as silent repairs. The same role-binding
and numerical checks are applied independently to $A$, $S$ and $D$.

The bounded contract fragment is a disjunction of conjunctions of numerical
comparisons over named roles, constants, absolute values and addition.
Configuration supplies RDF vocabulary, bounded focus-relative paths,
observation/subject types, metric identifiers and optional window/unit
requirements; predicate evaluation contains no financial rule-ID branches.
Unsupported expressions and exceeded assignment budgets raise errors rather
than produce analytical verdicts. This is an implementation fragment, not a
new expressive formalism or automatic translation of natural-language CQs.

For pool $P$, let $C_j(P)$ be the complete compatible assignments for branch
$j$, and $W_j(P)\subseteq C_j(P)$ those satisfying its comparisons. The audit
reports positive if some $W_j(P)$ is nonempty; false if every $C_j(P)$ is
nonempty and every $W_j(P)$ is empty; otherwise it reports unknown.
False therefore describes the inspected finite candidate pool under an
explicit complete-input policy, not an open-world refutation or a general
negation semantics. A passing OR branch suffices without the other branch;
missing evidence is never replaced with zero. No globally minimal witness or
complete failure proof is claimed.

Numerical status and declared-unit readiness are separate outputs; matching
unit labels do not perform conversion or authenticate a measurement. The
positive-readiness check requires all unit declarations on one and the same
passing assignment: a qualified nonpassing alternative cannot supply the units
of an unqualified passing witness. For a false diagnostic, readiness requires
a complete unit-ready assignment for each branch; this remains a finite-pool
policy, not a proof over unknown candidates. The
observation pools restrict membership, while entity/path/window metadata is
read from the fixed ambient snapshot. Consequently this audit checks delivered
observation sufficiency relative to that metadata, not whether an exported
standalone RDF subgraph is self-sufficient. Local graph/contract/checker hashes
identify experiment artifacts, not signed certificates or general blank-node
canonicalizations. These explicit boundaries make the audit testable against
an equivalent SHACL-SPARQL encoding without claiming greater expressiveness.

\hypertarget{the-ontokg-eq-method-and-system-architecture}{%
\section{The OntoKG-EQ method and system
architecture}\label{the-ontokg-eq-method-and-system-architecture}}

\begin{figure*}[t]
\centering
\includegraphics[width=0.85\textwidth]{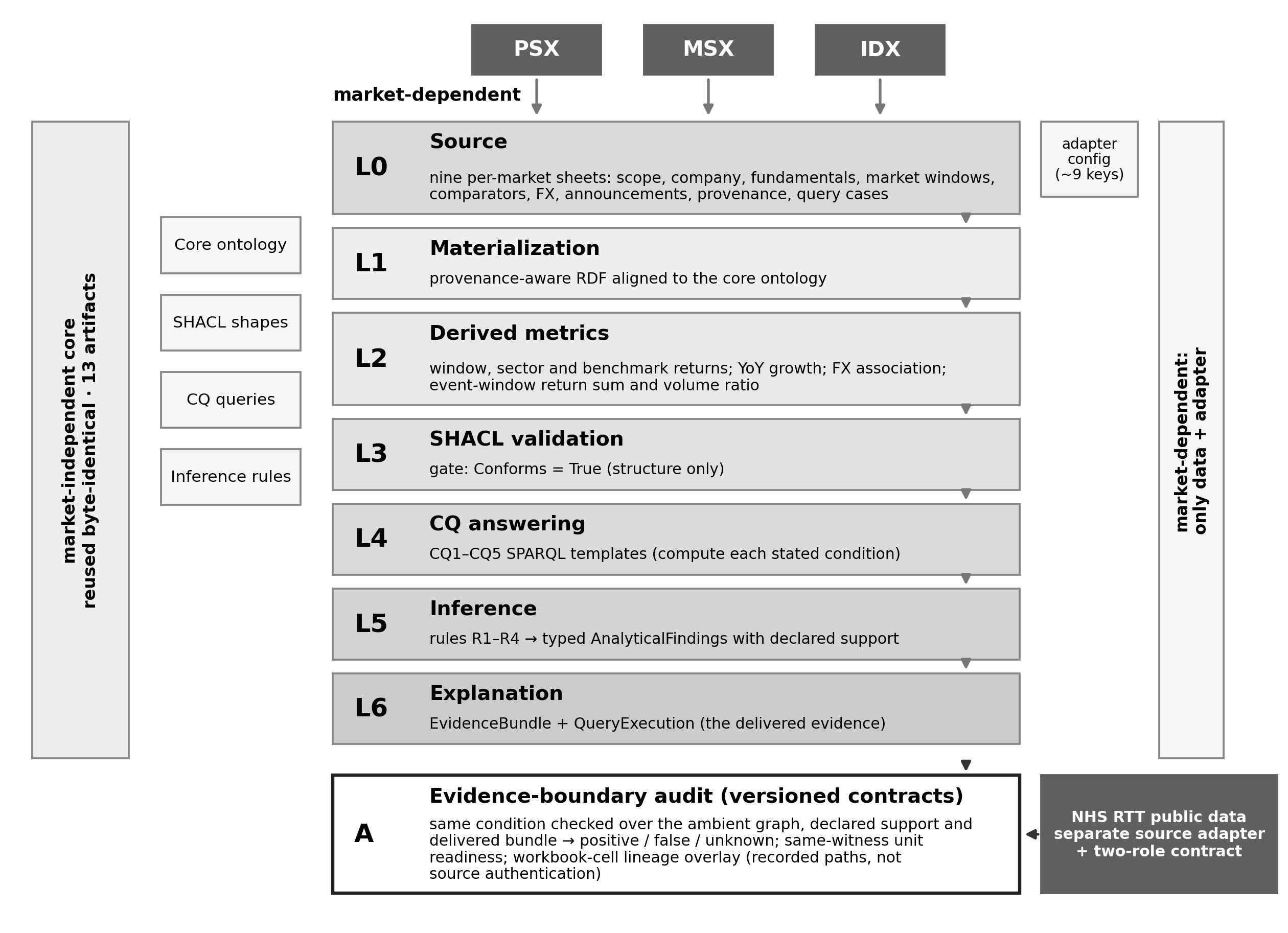}
\caption{The OntoKG-EQ pipeline. Layers L0--L6 build, validate, query and explain each market graph; only the L0 data and a small per-market adapter vary, while the ontology, shapes, CQ queries and rules are reused byte-identical. The evidence-boundary audit (A) checks findings against versioned contracts and is also applied to the NHS public-data application through a separate source adapter and contract.}\label{fig:architecture}
\end{figure*}

OntoKG-EQ is organised as a seven-layer pipeline followed by an audit layer (\hyperref[fig:architecture]{Fig.~\ref*{fig:architecture}}).

Layers L0--L1 onboard a market's data; L2 computes the analytical
metrics; L3 validates; L4 answers the competency questions; L5 derives
findings; and L6 generates explanations. The evidence-boundary audit (A; Sections~\ref{executable-contracts}--\ref{boundary-method}) then checks each finding's condition over the ambient graph, its declared support and its delivered bundle. The ontology, shapes, queries,
and rules (L1--L6 logic) are market-independent; only the data (L0) and
a small adapter configuration vary between markets.

\hypertarget{cq-bounded-core-ontology}{%
\subsection{CQ-bounded core ontology}\label{cq-bounded-core-ontology}}

The core ontology provides a compact vocabulary sufficient to express
the five competency questions and no more. Its principal classes are
Company, ObservedEntity, Observation (with subclasses
FundamentalObservation, MarketObservation, ExchangeRateObservation),
ReportingPeriod, AnalysisWindow, EventWindow, IndustrySectorClassifier,
IndustrySectorClassificationScheme, MarketIndex, Currency,
Announcement/Disclosure (as EvidenceItem), Publisher, EvidenceSource,
ProvenanceRecord, ValidationStatus, EvidenceBundle, and QueryExecution.
Object and datatype properties connect observations to their entities,
temporal contexts, metric values, evidence, and provenance. Each term is
traceable to at least one CQ (Section~\ref{competency-question-coverage-and-governance}), and the ontology classifies
without inconsistency under an OWL 2 EL reasoner. A small alignment
module maps the core to standard upper concepts where a CQ requires it,
keeping the core itself compact.

\hypertarget{l0l1-per-market-data-onboarding-and-materialization}{%
\subsection{L0--L1: Per-market data onboarding and
materialization}\label{l0l1-per-market-data-onboarding-and-materialization}}

A market is supplied as nine tabular sheets (scope/selection, company
master, fundamentals, market windows, comparators, exchange rates,
announcements/disclosures, provenance, and query cases), each row
carrying its source. The materializer reads these sheets and emits a
provenance-aware RDF graph aligned to the core ontology: companies and
their sector classifiers; reporting periods and analysis/event windows;
fundamental, market, comparator, and exchange-rate observations;
announcements as evidence items linked to publishers and evidence
sources; and provenance records. To remain reasoner-free at query time,
parent-class type assertions are materialized explicitly, so SPARQL and
SHACL operate without an external reasoner.

\hypertarget{l2-derived-analytical-metrics}{%
\subsection{L2: Derived analytical
metrics}\label{l2-derived-analytical-metrics}}

L2 computes the nine derived metrics of Section~\ref{core-constructs} as first-class
observations. The released builder computes the arithmetic sum of daily simple returns, multiplied by 100, over each company's post-report window when daily returns are available. This is a cumulative simple-return statistic, not a compounded holding-period return; when daily returns are unavailable, the builder uses the endpoint price return. These conventions must be distinguished when interpreting the demonstrator metrics; sector and benchmark returns are computed over the same window
dates as the company they are compared against, so that CQ1 and CQ3
compare like with like; exchange-rate association is the Pearson
correlation of daily exchange-rate returns with daily company and
benchmark returns over the shared window (of length n days; because such
short-window correlations are sensitive to outliers and
non-stationarity, the \textbar r\textbar{} ≥ 0.3 flag is a descriptive
screen, and Spearman and multi-window robustness checks are recommended
on licensed data, Section~\ref{threats-to-validity}), alongside the window-level exchange-rate
change. The event statistic currently labelled \emph{cumulative abnormal return} in the archived vocabulary is implemented as the arithmetic sum of supplied event \texttt{daily\_return} cells, multiplied by 100. The builder does not subtract benchmark returns at this step. Whether these cells already contain abnormal rather than ordinary returns requires upstream verification. Accordingly, we interpret the archived result as an event-window input-return statistic and do not claim that the current implementation independently performs a market-adjusted event study. The separate abnormal-volume ratio divides mean nonzero event volume by mean nonzero post-report volume. The lineage overlay records the actual cells used by these computations and leaves missing-value defaults unresolved.

For IDX, the acquisition assembler computes ordinary company returns as
consecutive close-price ratios and writes them unchanged to event rows.
Thus the IDX archived event statistic is a cumulative company-return sum,
not market-adjusted CAR. A market-adjusted event study would require aligned
daily company and benchmark returns with explicit subtraction. A market-model
extension would additionally estimate parameters over a prior window
\cite{ref35,ref36,ref37}. Neither refinement should be inferred from the
current graph's metric label.

\hypertarget{l3-structural-validation}{%
\subsection{L3: Structural validation}\label{l3-structural-validation}}

SHACL shapes encode the structural contract implied by the ontology:
e.g., every Observation carries exactly one metric name, one decimal
metric value, and at least one temporal context; a
FundamentalObservation is of a Company and observed in a
ReportingPeriod; an ExchangeRateObservation has base and dealt
currencies; an EvidenceBundle explains a result entity and includes at
least one observation or evidence item; and a QueryExecution has a
family identifier, an instance identifier, and an evidence bundle.
Validation is a gate: a market is only admitted to querying once its
graph reports Conforms = True.

\hypertarget{l4-answering-the-competency-questions}{%
\subsection{L4: Answering the competency
questions}\label{l4-answering-the-competency-questions}}

Each competency question is a parameterised SPARQL template that
\emph{computes its stated condition} rather than merely retrieving rows.
CQ1 selects companies whose year-on-year profit growth is positive but
whose post-report window return is below the benchmark return over the
same window; CQ2 returns the exchange-rate-versus-company and
exchange-rate-versus-benchmark return correlations together with the
window-level exchange-rate change; CQ3 selects companies whose window
return exceeds both their sector and benchmark returns over the same
window (joined on a shared window so the comparison is
period-consistent); CQ4 returns announcements whose event-window
cumulative abnormal return exceeds a threshold in magnitude or whose
abnormal volume ratio exceeds a threshold; and CQ5 assembles, for an
executed result, the supporting observations with their metric values,
the evidence items, and their official sources. The same five templates
run unchanged on every market.

\hypertarget{l5-rule-based-inference}{%
\subsection{L5: Rule-based inference}\label{l5-rule-based-inference}}

L5 applies SPARQL CONSTRUCT rules that \emph{materialise} each CQ
condition as an explicit, queryable typed AnalyticalFinding (Table~\ref{tab:t2}).
This is convenience packaging rather than a novel inference mechanism:
each rule evaluates a deterministic predicate over the derived metrics
(no rule interaction, conflict resolution, or learned parameter), whose
contribution is that a boolean condition becomes a first-class,
provenanced object rather than a transient query row. A finding records
the result entity, its type, the rule and CQ that produced it, and the
observations that triggered it, so the derivation itself is provenanced
and explainable. The event-window thresholds used by R4
(\textbar cumulative abnormal return\textbar{} ≥ 2\%, abnormal volume
ratio ≥ 1.5) instantiate an operational event-study screen (fixed
thresholds we do not tune) following \cite{ref35,ref36} and are fixed a priori
rather than tuned to the data.

\begin{table}[htbp]
\caption{Rule layer (L5): deterministic materialisation of each CQ condition as a typed, queryable AnalyticalFinding (not a learned or interacting rule system).}\label{tab:t2}
\centering\footnotesize\setlength{\tabcolsep}{4pt}
\begin{tabularx}{\linewidth}{llXX}
\toprule
Rule & CQ & Finding type & Condition \\
\midrule
R1 & CQ1 & fundamentals–market divergence & YoY profit growth > 0 and company window return < benchmark return \\
R2 & CQ2 & FX-sensitive & abs(FX-vs-company return correlation) ≥ 0.3 \\
R3 & CQ3 & relative-outperformer & company window return > sector and > benchmark return \\
R4 & CQ4 & abnormal-event reaction & abs(cumulative abnormal return) ≥ 2\% or abnormal volume ratio ≥ 1.5 \\
\bottomrule
\end{tabularx}
\end{table}

\hypertarget{l6-automated-explanation-generation}{%
\subsection{L6: Automated explanation
generation}\label{l6-automated-explanation-generation}}

For every materialized finding, L6 generates an EvidenceBundle and QueryExecution that link its result entity to the observations explicitly supported by the rule. All matching company announcements are retained as contextual evidence, instead of selecting an arbitrary last announcement. These announcement links alone do not establish which source supplied each metric.

The revised pipeline checks the final graph after adding findings, bundles, and executions, in addition to the initial structural gate. It also audits whether explicit finding supports and generated bundle supports agree and satisfy the stated rule condition. These checks are distinct from observation-level source attribution. In particular, the original R4 rule could select a finding through abnormal volume while attaching only the return observation; the revised rule includes the available volume observation. The final-graph audit in Section~\ref{jws-revision-audit} reports this correction and the remaining attribution boundary.

\hypertarget{the-portable-construction-methodology}{%
\subsection{The portable construction
methodology}\label{the-portable-construction-methodology}}

The pipeline doubles as a reusable methodology for onboarding a new
market: (Step 0) freeze the CQs; (Step 1) fix the market-independent
core (ontology, shapes, CQ templates, inference rules); (Step 2) acquire
the market's nine sheets, each row sourced; (Step 3) provide a small
\textbf{adapter configuration} (namespace, identifier and metric column
names, exchange-rate field, local currency, a growth-units scale, and
the benchmark label); (Step 4) materialize and derive; (Step 5)
validate, answer, infer, and explain; (Step 6) quantify portability.
Only Steps 2--3 are market-specific. The \emph{portability invariant} is
that, across markets, the ontology, shapes, queries, and rules are
byte-identical and only the data and the adapter differ, a property we
measure directly in Section~\ref{portability-across-three-markets}.

\hypertarget{implementation-and-reproducibility}{%
\section{Implementation and
reproducibility}\label{implementation-and-reproducibility}}

\hypertarget{software-components}{%
\subsection{Software components}\label{software-components}}

OntoKG-EQ is implemented in Python with a small, standard semantic-web
stack. The core ontology, alignment module, and demonstrator data are
authored in Turtle. SHACL shapes are validated with pySHACL. The RDF
graph is built and queried with RDFLib. The nine input sheets are read
with openpyxl. The pipeline is realised by three components that
correspond to the layers of Section~\ref{the-ontokg-eq-method-and-system-architecture}: a unified builder that performs
materialization (L1), derived-metric computation (L2), and assembly of
the worked CQ instances; the SHACL shapes and CQ templates that drive
validation (L3) and answering (L4); and an inference-and-explanation
generator that applies the rules (L5) and auto-generates the evidence
bundles and query executions (L6). The inference rules are additionally
published as standalone SPARQL CONSTRUCT files and a small findings
vocabulary, so the reasoning layer is inspectable independently of the
driver code.

\hypertarget{one-pipeline-a-per-market-adapter}{%
\subsection{One pipeline, a per-market
adapter}\label{one-pipeline-a-per-market-adapter}}

The builder contains a single code path; each market is described by one
configuration record (Supplementary Table S1), the only thing that
changes between markets apart from the data. A new market is onboarded
by adding one such record and supplying its nine sheets.

\hypertarget{recorded-environment-and-availability}{%
\subsection{Recorded environment and
availability}\label{recorded-environment-and-availability}}

The original deterministic results were produced with Python 3.10 and pinned
versions of RDFLib, pySHACL, openpyxl, pandas, and pyoxigraph (Oxigraph)
on a single workstation (2 vCPU, 3.8 GB RAM); the exact package versions
and hardware are recorded in \path{software_environment.md}; core
dependencies are specified in \path{requirements.txt} and exactly
pinned in \path{requirements-lock.txt}. The eight-model LLM
faithfulness panel was run separately on a Kaggle GPU. Its notebook pins
the package versions in the install cell and records the exact resolved
versions, CUDA/GPU, and Hugging Face model revisions in
\path{environment_lock.json}. The pipeline is deterministic:
rebuilding a market from its sheets reproduces the same graph, SHACL
verdict, and competency-question results. The complete reproducibility
package (ontology, alignment, SHACL shapes, the five CQ templates, the
inference rules and findings vocabulary, the unified builder and the
inference/explanation generator, the per-market data and materialized
graphs for all three markets, the evaluation scripts and outputs, a data
dictionary, and the recorded environment) is openly released in a public
repository with citable Zenodo archives (original release \url{https://doi.org/10.5281/zenodo.21569316}; revision release v2.0.0 \url{https://doi.org/10.5281/zenodo.23860671}). For Indonesia, the exact
data-acquisition scripts, the strict exact-date intersection builder,
and a source manifest are included. Earlier revision checks used Windows, Python 3.13,
RDFLib 7.6.0, pySHACL 0.31.0 and openpyxl 3.1.5. Final local verification used
Python 3.13.5, RDFLib 7.6.0, pySHACL 0.40.1, openpyxl 3.1.5 and Oxigraph 0.5.11;
the separate revision dependency file and run ledger record this environment.
Script outputs record graph hashes and coverage. Historical performance
measurements were not rerun in this environment. Participant-level study data
remain private under the original consent terms, and historical model answer
texts were not archived. These limitations constrain independent reproduction.

\hypertarget{experimental-setup}{%
\section{Experimental setup}\label{experimental-setup}}

\hypertarget{markets-and-companies}{%
\subsection{Markets and companies}\label{markets-and-companies}}

We evaluate on three independent emerging equity markets, each onboarded
through the nine-sheet process of Section~\ref{the-portable-construction-methodology} with two sectors and two
companies per sector (Supplementary Table S2). The two-sector,
two-company design keeps the slice small enough to audit end to end
while still exercising sector peer-baskets (needed by CQ3) and
cross-sector comparison.

\hypertarget{data-sources-and-provenance}{%
\subsection{Data sources and
provenance}\label{data-sources-and-provenance}}

Each market integrates company fundamentals (annual EPS and year-on-year
profit growth), daily market windows (close and volume around the
reporting anchor), sector peer-basket and broad-market benchmark
comparators, a daily local-currency/USD exchange-rate series, and
official annual-report announcements with disclosure references.
Fundamentals come from public annual-report summaries, market and
exchange-rate series from public historical price sources, and
announcement anchors from official exchange disclosures; benchmark and
sector-basket comparators are transparent demonstrator constructions. PSX benchmark rows declare interpolation/forward-filling, and MSX benchmark rows declare a damped selected-universe proxy, not an official KLCI series. Their findings test stored-data semantics rather than empirical market reactions.
Every row records its source, retrieval date, and field, carried into
the RDF provenance layer. Indonesia is the strongest provenance case:
its six daily series were downloaded from a public endpoint and
\emph{strictly intersected on exact trading dates} (no interpolation,
forward-fill, or synthetic rows), yielding 83 common trading dates
spanning 2026-01-15 to 2026-05-29; one telecom constituent (TLKM) was
replaced by TOWR because its FY2025 results were released too late for a
completed post-event window, a decision recorded in the bundle.

\hypertarget{tasks-metrics-and-baselines}{%
\subsection{Tasks, metrics, and
baselines}\label{tasks-metrics-and-baselines}}

The original financial tasks are the five competency questions of
Section~\ref{frozen-competency-questions}, executed on each market. The
additional public-data task applies the evidence-boundary interface to a
provider-month statistical screen rather than reusing the financial ontology.
It selects all eligible published August 2026 records, not a sample chosen
for positive findings. A source oracle and a separately written source adapter
parse the original CSV and workbook independently; neither imports the audit
interpreter. Exact aggregate counts, reporting month and source selectors are
compared before the generated RDF views are audited. The matched comparator
executes an independently authored SHACL-SPARQL constraint and availability/unit
queries on the same evidence boundaries. Primary outcomes are source
reconciliation, numerical status, same-witness unit readiness and baseline
agreement, not clinical accuracy, human task speed or causal impact.
Section~\ref{nhs-application} reports population, missingness and adaptations.
We report the source-reconciled application and corrected contract evaluation first,
followed by archived coverage, portability, relational and structural controls,
ablations, scalability and the historical perception study
(Section~\ref{explanation-quality}). Archived execution results refer to their
recorded environment. Private participant responses and missing historical
model-answer texts limit independent reproduction; they are not claimed to be
reconstructible from the public bundle.

\hypertarget{results-and-evaluation}{%
\section{Results and evaluation}\label{results-and-evaluation}}

\subsection{Strengthened evaluation: corrected metrics and explicit failure cases}\label{strengthened-evaluation}
\paragraph{Genuine public-data application: NHS waiting-time statistics.}\label{nhs-application}
The nonfinancial application reconstructs NHS England's August 2026
provider-level incomplete referral-to-treatment (RTT) screen from its original
CSV archive and provider workbook~\cite{nhsRTTRelease,nhsRTTGuidance}.
The publisher-defined question is whether at least 92\% of incomplete pathways
are within 18 weeks, not a hospital-quality ranking or a clinical decision.
The numerator sums eighteen weekly bands, 0--1 through 17--18 weeks;
the denominator is \texttt{Total All}. The exact predicate is
$T>0\;\land\;100N\geq92T$, avoiding rounded-percentage threshold decisions.
All \path{Part_2}, all-specialty rows are included except commissioner
\texttt{NONC}; commissioner rows are aggregated once per provider and month.
A dated protocol preceded provider-outcome inspection, but is not preregistration.

A source oracle independent of the RDF adapter and checker reconciles all
532 eligible providers against the publisher workbook: eighteen weekly counts,
the numerator and denominator agree exactly; stored ratios agree within the
fixed $10^{-12}$ tolerance in ratio units. Exact cross-products classify
91 providers at or above 92\% and 441 below. The workbook roster contains
one additional provider, \texttt{D1V5N}, with zero published denominator but
no selected CSV records. Its CSV-derived counts remain unavailable, rather
than being zero-filled or interpreted as a failed target or missing submission.
The primary units are 532 eligible provider-month records and one unavailable
roster record, not three independent observations per evidence pool.

The application-specific engineering is explicit: a new source adapter,
a two-role RDF vocabulary and contract, and bounded fixed-integer scaling
compiled into balanced repeated addition. The existing bounded interpreter
is unchanged; no multiplication operator or financial ontology reuse is claimed.
Aggregation of the eighteen input bands occurs in the recorded source adapter,
not inside the two-role interpreter. Generated evidence views contain the
deciding counts and selectors for original CSV rows and columns, anchored by
source-file hashes. These are our outputs over genuine source records, not
publisher-provided explanation bundles. The recorded selectors reference
242,934 source cells; this is neither that many independently collected
observations nor authenticated patient-level lineage.

All 533 application
frames agree with an independently authored, executed SHACL-SPARQL and explicit
availability/unit-readiness baseline on all three pools: 91 are positive,
441 false and one unknown. Complete generated views deliberately carry the same
deciding observations across the three pools, so these are 533 matched frames,
not 1,599 independent successes. The baseline reuses identical within-frame
pool results and executes 533 actual SHACL validations. Its SPARQL arithmetic
uses explicitly sufficient Decimal precision; this is disclosed comparator
configuration, not a stronger SHACL semantics. The generated dataset contains
18,095 RDF triples. These measurements show successful technical execution and
traceable deciding evidence, not superiority to the matched baseline.

This establishes a reproducible externally specified statistical task and
source-to-evidence traceability beyond financial fixtures. The CSV and workbook
share a reporting collection, so their reconciliation is not independent data
collection. No natural reconciliation defect, independent production deployment,
clinical benefit or improvement in human audit performance is established.
Information reuse follows the publisher's OGL terms and exceptions; original
branded binaries are retained locally, not redistributed as an endorsed NHS
product~\cite{nhsTerms,ogl3}.

\paragraph{Requirement changes on the public-data application.}\label{nhs-change}
A separately recorded protocol replays six threshold/operator policies on the
same 533 NHS application frames. Only the inclusive 92\% screen is official;
all alternatives are counterfactual requirements, not changes to NHS policy
or reporting errors. An exact-integer oracle computes expectations from the
existing adapter records. An independently parameterized direct comparator
executes prepared SPARQL queries with provider bindings. Both implementations
centralize policy parameters; neither is forced to duplicate manual edits.
All provider identities, numerical statuses, unit-readiness values and exact
affected-provider sets agree. Table~\ref{tab:nhs-change} reports sensitivity;
all policies preserve the same unavailable roster entry.

\begin{table}[htbp]
\centering\small
\caption{Policy replay on the same 533 frames. Only the inclusive 92\% rule is
official. Changes are counterfactual reclassifications, not source errors.}
\label{tab:nhs-change}
\begin{tabular}{lrr}
\toprule
Policy & Positive & Changed from official \\
\midrule
Official $\geq92\%$ & 91 & 0 \\
Counterfactual $\geq90\%$ & 118 & 27 \\
Counterfactual $>90\%$ & 117 & 26 \\
Counterfactual $>92\%$ & 91 & 0 \\
Counterfactual $\geq95\%$ & 67 & 24 \\
Counterfactual $>95\%$ & 67 & 24 \\
\bottomrule
\end{tabular}
\end{table}

Fourteen constructed controls per policy exercise equality, missing evidence,
shared months, delivered omission and same-witness units, including 252 actual
SHACL pool validations; cohort SPARQL execution is not counted as SHACL
validation, and repeated policies do not enlarge the independent sample.
Re-executing the official policy restores every status/readiness result, and
source-derived records, earlier outputs and implementation files retain their
recorded hashes with no core-interpreter edit. This is deterministic replay,
not stateful deployment rollback. An initially unbound query attempt was
interrupted before completing a policy; its retained note documents the switch
to provider-bound evaluation of the same query texts. The experiment
establishes bounded change-conformance and affected-result traceability. The
direct implementation succeeds equally, so reduced authoring effort, superior
maintainability and practitioner benefit do not follow.

\paragraph{Declarative evidence-boundary diagnostics.}
The additive interpreter of Section~\ref{boundary-method} executes twenty
constructed sensor cases with independently specified ambient, support and
delivered labels. All sixty pool-level results agree with those expectations
and with a separately authored SHACL-SPARQL positive-witness constraint plus
availability queries. The comparator independently reconstructs pool membership
from the fixture RDF; it does not reuse the interpreter's selections.
Its coverage is the sensor OR configuration and ordinary-decimal lexical
inputs, not all financial predicates, metadata readiness or arbitrary RDF
numeric terms. Agreement establishes functional parity on this diagnostic
subset, not better accuracy or expressiveness than SHACL-SPARQL.

A second, independently specified analytical suite closes part of that
coverage gap. Twenty-six constructed cases yield 78 paired pool comparisons
with no numerical-status or unit-readiness disagreement against explicit
expectations and manually authored SHACL-SPARQL/availability queries.
All four financial finding families are exercised through the full audit
entry point, including conjunctions, absolute values, addition, shared-window
joins, typed role/path decoys and invalid bundle associations. R3 uses an
explicit diagnostic margin of one percentage point without modifying the
frozen zero-margin configuration. A passing but unit-unqualified witness
cannot borrow units from a qualified nonpassing alternative; adding a
qualified passing alternative does establish readiness. Three regression
methods check the suite, a wrong-label failure guard and this witness-coherence
case. The comparator independently selects its observation pools from RDF.

Primary parity is restricted to the tested short-precision decimal fixtures
(at most five significant digits and four decimal places), not general
numeric equivalence. A separately retained precision stress case lies outside
that subset: with sector $2.00000000000000000000000000001$, company
$3.00000000000000000000000000001$ and margin one, exact replay is false but
the default-precision RDFLib SPARQL comparison is positive due to addition
rounding. This mismatch is recorded separately, not silently discarded or
counted among the 78 agreements. Increased comparator precision can address
this example; it is not a SHACL expressiveness limitation. The expanded
diagnostics remain constructed workload evidence, not a new empirical sample
or a proof of conceptual novelty. Queries, parameters, hashes and outcomes are
released separately in \path{boundary_analytical_v1/}.

\begin{table}[htbp]
\centering\small
\caption{Selected constructed evidence-boundary cases. These are examples from
twenty diagnostics, not observed failure frequencies; metadata remains ambient.}
\label{tab:boundary-cases}
\begin{tabularx}{\linewidth}{Xlll}
\toprule
Controlled change & Ambient & Support & Delivered \\
\midrule
All deciding observations included & Positive & Positive & Positive \\
Remove delivered observations only & Positive & Positive & Unknown \\
Deliver only nonpassing candidates & Positive & Positive & False \\
Remove declared support only & Positive & Unknown & Unknown \\
\bottomrule
\end{tabularx}
\end{table}

Seven additional regression methods exercise the diagnostic suite, unit
readiness, bundle association, invalid configurations, enumeration bounds,
all four financial configurations, incompatible joint windows and exact
addition/absolute value under reduced ambient Decimal precision. These tests
are not an external comparator for every supported expression. The same
interpreter consumes financial and synthetic sensor vocabularies without
code changes: that synthetic experiment alone establishes configuration-level
reuse, not real-domain validation; the NHS application is reported separately.
Single-run diagnostic timings include comparator
setup and are not used to claim a speed advantage.

A separate paired replay evaluates the frozen financial experiment pools
with the declarative numerical interpreter. All 487 existing records agree
on support/delivered status and ambient truth. This comparison uses direct
pool evaluation, explicit in-memory company-to-index lookup adapters and
the frozen R4 event scope; it does not test new bundle-link selection,
metadata readiness or source-path parity. Input hashes remain unchanged.
The records are overlapping archived/corrected experiment versions, not an
additional independent sample. Diagnostic and paired results are recorded
separately in \path{boundary_audit_v1/} and
\path{boundary_legacy_parity_v1/}.

\paragraph{Actual stored-release evidence omissions.}
The audit entry point is also run on the actual original and repaired
three-market releases, not only on constructed findings. Original bytes
retrieved from a fixed Git commit and repaired files match their preceding
release-audit hash ledgers. Explicit in-memory adapters link findings to
already existing generated bundles and expose the archived implicit index
lookup; no numerical value, support link or bundle member is repaired by the
adapter. Support and delivered sufficiency are 19/21 in the original release
and 21/21 after repair. Exactly the two previously documented PSX R4 volume
omissions change from unknown to positive. These are 21 paired findings,
not 42 independent examples, and the experiment requires the recorded source
history rather than the source-only review ZIP alone. Original R4 event
identity is insufficient for independently anchored ambient replay, so ambient
positives are not treated as event-specific truth guarantees. The synthetic
matched baseline is a separate experiment, not a comparator run on these
release findings. Results and adapter details are recorded in
\path{boundary_release_replay_v1/}.

\paragraph{Financial fixture recomputation and qualification.}
The archived results reported later remain historical controls. A separate financial-v2 release recomputes company, sector and benchmark price returns by compounding consecutive close-price intervals over exactly the same company dates. Missing comparator endpoints produce an incomplete result, not a silently shortened window. These are price returns, not authenticated dividend-adjusted total returns. Event CAR is now explicitly the sum of supplied company daily returns minus same-date benchmark returns, in percentage points; missing operands are incomplete. Volume ratios retain legitimate zeros and disclose baseline overlap. FX association uses identical intervals with at least three pairs. Growth uses matching current/prior profit fields where available, with an absolute prior-value denominator; supplied growth otherwise remains explicitly identified. Currency context is separate from metric units. This revision does not change archived workbooks or graphs.

Across four corrected datasets, 404 metrics were recomputed and 17 marked incomplete; 5,237 supplied daily-return/adjacent-price checks agree. A bounded primary-source spot audit found monetary scaling errors in ENGROH and FFC workbook fields. Eight reviewed cells now carry separately recorded normalized values, multipliers and source citations; the raw cells remain unchanged. Common scaling may cancel in a growth ratio but does not make the monetary extraction accurate. This audit does not authenticate all prices, financial facts, dates or upstream derivations.

Expanded evaluation enumerates all company classification cases rather than inspecting positives alone. The corrected 64-company cohort contains 42 numerical positives, 147 negatives and three incomplete cases for R1--R3; its 64 event classifications are separately not applicable because event observations are absent. The corrected curated cohorts contain 16 positives. Curated IDX overlaps the scaled cohort, and original/corrected datasets are paired, so these counts must not be summed as independent samples. R2 remains a classification test, not validation of the entire CQ2 answer.

Original unbound CQ1/CQ3 queries evaluated on complete graphs agree with an independent Decimal traversal for all 16 dataset/query result sets. Separate experiment findings and actual bundles replay all enumerated classifications. Removing decisive evidence yields the expected response in 122/122 corrected scaled-cohort and 38/38 corrected curated-cohort variants; an independently sufficient OR branch remains valid. Eleven synthetic boundary/branch controls are reported separately from natural cases. These checks establish stored-data semantics, not external financial truth.

A subsequent categorical-accounting run preserves every input hash and numerical cohort count while checking status as well as binary sufficiency. All 487 replay records agree on positive/sufficient/true, negative/predicate-false/false or incomplete/incomplete-bundle/unknown status, under the explicitly tested R4 missing-branch convention. All sixteen complete-query comparisons and 330 withholding variants across the paired cohorts agree. These totals include overlapping versions and constructed experiment bundles, not 487 independently collected findings or 330 natural observations. Seventeen synthetic graph configurations separately exercise categorical and branch behavior. Frozen v1 artifacts remain unchanged; the stricter run is recorded in \path{expanded_evaluation_v2/}.

An independent metadata audit checks role-specific units, comparator currency contexts, R1 report-to-window linkage and R4 explicit event-window linkage. Only 22/42 scaled positives and 15/16 curated positives satisfy all applicable declarations. Twenty scaled R1 cases use a recent-analysis anchor different from their recorded report date. Curated TOWR records 18 March 2026 as the report date but 25 March as the window anchor. They remain numerical positives, not validated report-anchored findings. Dates are not rewritten to manufacture agreement. Five synthetic metadata controls pass; archived cases without required declarations remain unverified. Declaration consistency is weaker than date authenticity or knowledge of announcement time.

\paragraph{Source qualification beyond arithmetic.}
Stored-data replay, matching declarations and authentic source inputs are separate claims. Versioned repairs now authenticate twelve OGDC/PPL annual facts: eight monetary cells agree only after a factor of one million corrects their declared thousand-PKR scale, while four EPS cells match directly. The original wrong-period links and raw workbooks remain preserved. FY2024 values are checked as FY2025 comparative columns, not against separate original releases. Company/separate and unconsolidated accounting scopes are retained. Growth ratios remain unchanged because common scaling cancels, but this does not validate monetary unit labels.

\begin{table}[htbp]
\centering
\caption{Current source-backed evidence and its limits. Counts describe different units and overlapping subsets; they are not an independent sample total.}
\label{tab:source-qualification}
\small
\begin{tabularx}{\linewidth}{p{0.21\linewidth}XX}
\toprule
Evidence & Executed coverage & Remaining qualification \\
\midrule
OGDC/PPL annual facts & 12 facts across two companies and two year columns & Eight require scale correction; scope and comparative-year limits remain \\
Maybank company closes & 32 required dates present; all 30 archived post-report closes match & Issuer republication; adjustment convention unresolved \\
SECP PSX reports & Four original reports; six KSE100 levels and two FFC closes & KSE100 total return is not a price-only benchmark; FFC baseline missing \\
Disclosure dates & Two official dates match FFC/Maybank anchors & Document/annual-report dates do not establish first earnings-publication times \\
Source-backed comparisons & Two existing cohort subsets replayed; zero eligible market-relative comparisons & Missing compatible benchmarks, definitions and baseline coverage; labels withheld \\
Frozen source sample & 52 cases; 2,357 unique recorded workbook-cell inputs & Zero cases have complete reviewed coverage under the all-recorded-input policy \\
\bottomrule
\end{tabularx}
\end{table}

The supplied PSX comparator declares interpolation and forward-filling, and MSX declares a damped selected-universe proxy rather than official KLCI data. Their classifications are illustrative predicate tests, not authenticated market reactions. Authentic KSE100 total-return levels cannot substitute for KSE100PR in a company price-return comparison. A PSX company-page definition applies only to that feed, not automatically to original SECP closes or other providers. Maybank's unexplained \path{unadjusted=0} parameter is not treated as an adjustment definition. Corporate-action review remains incomplete.

\paragraph{Systematic dependency coverage.}
An executed audit of the frozen 52-case, 29-stratum frame enumerates 2,357 unique recorded workbook-cell inputs. Of these, 2,271 have recorded lineage but no independent review, 46 are declared proxies, and two reconstructed KSE100 values disagree with original regulator observations. The workbook records 157,496.1 for both 12 and 13 March 2026, whereas the original report's explicitly dated columns give 154,421.43 and 153,866.17. The remaining 38 inputs have different bounded qualifications: four exact scale-corrected PAT values, two rounded narrative agreements, thirty issuer-republished Maybank closes with unresolved adjustment basis, and two regulator numeric agreements without transfer of the workbook's basis. These are not equivalent authenticity statuses.

No sampled case has complete independently reviewed coverage under the conservative all-recorded-input policy. This is not a claim that all cases are false, nor a minimal sufficient-branch AND/OR proof or population accuracy estimate. Dependencies are deduplicated by exact workbook/hash/sheet/row/field identity; overlapping datasets and repeated cells are not independent observations. The six newly verified PPL facts fall outside this frozen dependency selection and are not added to its coverage count. OGDC review uses a separate repair overlay and does not silently fix historical graph source links. Thirteen assertions check the frozen frame, graph/source hashes, joins and coverage accounting. The result qualifies the paper as a semantic-contract and provenance evaluation on explicit financial fixtures, not a source-authenticated empirical market study.

\paragraph{Temporal eligibility and bounded recomputation.}
All 20 scaled R1 positives are quarantined from report-event use because their workbooks explicitly identify recent-analysis windows and use fiscal year-end as a purported report date. Curated TOWR's 18/25 March mismatch is also quarantined. The 22 retained scaled positives are non-R1 classifications, not authenticated earnings events. Among all eight PSX/MSX candidates, three numerical R1 positives and five negatives are retained in the ledger; only FFC and Maybank match official document/annual-report dates. Four lack matching verified dates and OGDC/PPL lack September 2025 price coverage for their corrected annual sources.

The two matched cases contain 30 post-disclosure closes and 29 observed intervals. Recovered company-only endpoints can support descriptive arithmetic, but not a complete initial-reaction baseline or a compatible market-relative result. The consolidated source-backed replay therefore leaves both market-relative labels unknown, rather than converting missing evidence into negative findings. These two cases overlap earlier cohorts and do not increase the independent sample. Detailed formulas, values, field/page locators, source hashes, acquisition failures and successive repair stages appear in the supplement and versioned ledgers. Original graphs remain unchanged; no market data are silently interpolated or substituted.

\subsection{Executed external provenance comparison}\label{npcs-comparison}
We executed the published, unmodified NPCS Standard-reification JAR at archived revision \path{b95bff3b56587b2a9ef6655503aada19c844630d}. Original and rewritten queries use the same Oxigraph engine. For 1,044 observation tuples, tuple multisets agree and each native provenance polynomial identifies exactly four expected supporting facts. For the full 15-pattern CQ3 basic graph pattern, all 12 candidates agree and each polynomial identifies exactly 15 facts. Explicit numerical post-filtering produces the archived PSX/MSX/IDX selections of 2/0/2; contract interpretation agrees on four positives and eight negatives.

The native numerical-FILTER rewrite was also tested: the emitted query double-reifies object patterns, omits the provenance projection and returns zero rows on these inputs. Generated queries are retained. Consequently this is a genuine external BGP provenance comparison with explicit analytical post-processing, not an unmodified end-to-end CQ3 benchmark or evidence of general superiority. NPCS provides query-support lineage; contracts add application-specific interpretation. The earlier constructed corruption suite remains a policy ablation and must not be attributed to NPCS. Single-run timings are descriptive, not a speed ranking.

\subsection{Objective human evaluation: prepared, not executed}\label{objective-study}
The historical 17-person study measures perceived trust and completeness only. A separate objective protocol now provides 12 distinct company/window audit tasks: four supported, four contradicted, and four with insufficient comparator evidence. Two displays contain identical facts and source appendices, differing in role organization; neither reveals a validator verdict. Four sequences counterbalance display and order. Each participant sees each case once, with six tasks per display and balanced truth categories. Outcomes are decision accuracy, source-identification accuracy and completion time, analyzed at participant level with paired differences and participant bootstrap intervals. Consent/ethics coverage, sample-size justification, allocation and any preregistration must be settled before recruitment. Packets and analysis software exist, but no new participants or responses have been generated. Objective audit improvement is therefore an open empirical question. Public provenance-interaction records can inform secondary analysis and task design but cannot estimate the effect of these displays without participants using them.

\hypertarget{competency-question-coverage-and-governance}{%
\subsection{Competency-question coverage and
governance}\label{competency-question-coverage-and-governance}}

We first verify the governance invariant. Introspecting the core
ontology yields 24 classes, 22 object properties, and 23 datatype
properties; together with the 9 derived metrics and 15 SHACL shapes,
\textbf{every term maps to at least one competency question, with no
uncovered terms} (the check is mechanical and ships in the bundle).
Conversely, each competency question is answerable end to end: executing
the five SPARQL templates returns results on at least one market for
every family. The ontology is thus both sufficient (answers all CQs) and
disciplined (no term without a CQ). We claim traceability and controlled
growth, not formal ontological minimality.

\hypertarget{portability-across-three-markets}{%
\subsection{Portability across three
markets}\label{portability-across-three-markets}}

Across PSX, MSX, and IDX, \textbf{13 core artifacts are reused
byte-identical}: the core ontology, project, and alignment files, the
SHACL shapes, the five CQ templates, the inference rules and findings
vocabulary, the unified builder, and the inference/explanation
generator. That is \textbf{100\% of the ontology, shapes, queries,
rules, and code logic}. Only the nine data sheets and a single adapter
record (Supplementary Table S1) differ per market. We are explicit about
what this does and does not show: each market's raw sources are first
\textbf{mapped by hand into the common nine-sheet schema}, and only then
is the apparatus reused unchanged. The 100\% figure therefore measures
reuse of the ontology, shapes, queries, and rules \emph{over an
already-aligned representation}, not automatic ingestion of arbitrary
raw data (onboarding a genuinely different reporting structure would
require new source mappings whose effort we have not measured; Section~\ref{limitations-and-future-work}). Table~\ref{tab:t5} reports the per-market outcomes under this identical
apparatus.

\begin{table*}[htbp]
\caption{Three markets, one unchanged pipeline. CQ columns report the number of result rows. The CQ2 count is \textit{structural} (one exchange-rate-association row per company by construction, ≈ one per firm), not a data-dependent outcome. CQ1/CQ3/CQ4 can legitimately return empty. All quoted returns are demonstrator quantities (Section~\ref{threats-to-validity}), and "benchmark return" is measured over each company's own window.}\label{tab:t5}
\centering\scriptsize\setlength{\tabcolsep}{4pt}
\begin{tabular}{lrcrrrrrr}
\toprule
Market & RDF triples & SHACL & CQ1 & CQ2 & CQ3 & CQ4 & CQ5 & Findings \\
\midrule
PSX (KSE-100) & 4,065 & Conforms & 0 & 4 & 2 & 6 & 3 & 9 \\
MSX (Bursa) & 2,618 & Conforms & 2 & 4 & 0 & 0 & 3 & 2 \\
IDX (Jakarta) & 4,185 & Conforms & 0 & 4 & 2 & 6 & 3 & 10 \\
\bottomrule
\end{tabular}
\end{table*}

All three markets conform to the same SHACL shapes and answer the
competency questions with traceable explanations. PSX and IDX, built
independently, happen to share the same CQ result \emph{shape}
(0/4/2/6/3). With only four firms per curated market, we read this as
\emph{illustrative} rather than strong statistical evidence, resting the
portability claim on the byte-identical apparatus (100\% reuse) and the
64-stock scaled run (Section~\ref{scaling-to-a-64-stock-cross-section}). The non-empty cells are real
analytical answers over demonstrator comparators (all quoted returns are
demonstrator quantities, not licensed estimates, Section~\ref{threats-to-validity}): for
example, CQ3 on PSX returns OGDC (+8.11\% \textgreater{} sector +7.28\%
\textgreater{} benchmark −9.11\%) and ENGROH, CQ1 on MSX returns MAXIS
(+11.82\% year-on-year profit growth but −6.31\% return versus a +0.16\%
benchmark), and CQ3 on IDX returns ISAT and BMRI. Empty cells are honest
data outcomes, not failures. In a falling Jakarta market the large caps
all beat their benchmark, so CQ1's set is empty, and different families
are non-empty across markets because the data differ while the method
does not.

\hypertarget{scaling-to-a-64-stock-cross-section}{%
\subsubsection{Scaling to a 64-stock
cross-section}\label{scaling-to-a-64-stock-cross-section}}

To show the method is not confined to small demonstrator slices, we
scaled the Indonesia market to a \textbf{64-stock cross-section spanning
11 GICS sectors}, acquired from public daily data over a common
2026-03-26 to 2026-06-29 window (real prices and volumes, the Jakarta
Composite, and USD/IDR, plus two years of fundamentals per stock for
year-on-year growth). The unchanged pipeline materialized a
37,046-triple graph that is SHACL-conformant, and the competency
questions return substantive cross-sections: 17 companies with positive
reported growth yet a post-window return below the −19.4\% Jakarta
Composite benchmark (CQ1, on the \emph{sign} of year-on-year growth at
this scale; see the caveat below); 24 companies that outperform both
their sector and the benchmark (CQ3, e.g.~MAPI +31.6\% vs sector +13.4\%
vs benchmark −19.4\%); 63 FX-association rows (one per company; one
stock lacked sufficient overlapping FX-return observations, so 63 of 64)
(CQ2); and the explainability query (CQ5). Event-window observations
(CQ4) are not generated at this scale because per-stock announcement
anchors were not collected, so CQ4 remains exercised on the curated
four-stock set. The 64-stock demonstration therefore exercises CQ1
(sign-restricted), CQ2, CQ3, and CQ5 at scale. The two multi-join
queries (CQ1, CQ3) exceed RDFLib's in-memory nested-loop engine at this
size (Section~\ref{scalability}). We therefore executed the \textbf{unchanged} SPARQL
templates on a standard triplestore (Oxigraph), where CQ1 returns its 17
rows in ≈4 ms and CQ3 its 24 rows in ≈5 ms over the full 37,046-triple
graph (medians; Supplementary Table S5) (the complete CQ1--CQ5
triplestore latencies are reported in Section~\ref{scalability}). These match the
counts from directly evaluating the same conditions over the derived
metrics, so the cross-sections come from the system's own query layer,
not a bypass. One data caveat: Yahoo-derived year-on-year net-income
growth can be extreme on small or recovering bases (a few firms exceed
several hundred percent), so CQ1 uses the \emph{sign} of growth rather
than its magnitude; representative moderate cases include CPIN (+52\%)
and JPFA (+33\%). This removes the small-demonstrator concern with the
ontology, shapes, queries, and rules byte-identical. The acquisition
script is released in the bundle.

\hypertarget{controlled-evidence-bundle-to-text-faithfulness-evaluation}{%
\subsection{Excluded historical measurements}\label{controlled-evidence-bundle-to-text-faithfulness-evaluation}}
The historical eight-model experiment supplied structured evidence to instruction-tuned language models and scored their textual transcription, with retrieval held fixed. The archived notebooks and score tables do not contain the generated answer texts needed to apply the current, differently defined scorer, so we exclude the historical model ranking and make no quantitative claim from it. Historical timings compared deterministic rendering of precomputed answers with model generation on different hardware; these are different tasks and establish no efficiency advantage. Query-layer performance is assessed in Section~\ref{scalability}. A reproducible future transcription experiment must archive each answer, input bundle, source identity, prompt, model revision and scorer version.

\hypertarget{ablations-and-the-ungoverned-knowledge-graph-baseline}{%
\subsection{Component-removal ablations}\label{ablations-and-the-ungoverned-knowledge-graph-baseline}}

Component removal tests characterize implementation dependencies. Removing evidence links reduces the historical evidence-availability score; removing rules removes the findings produced by those rules; and removing bundles removes bundle-query output. These outcomes establish that the implementation uses the removed components. They do not compare OntoKG-EQ with a capable provenance system or establish that relational stores cannot provide those functions.

Stronger final-graph interventions are reported in Section~\ref{jws-revision-audit}.

\hypertarget{scalability}{%
\subsection{Scalability}\label{scalability}}

Replicating a market graph with constant schema and multiplied
instances, SHACL validation scales near-linearly on RDFLib (≈0.11 s per
1k triples) and the single-pattern competency questions stay flat, but
the two multi-join analytical queries (CQ1, CQ3) degrade sharply under
RDFLib's in-memory nested-loop join (not completing within two minutes
on the 37k-triple graph). This is an \emph{engine} limitation, not a
method one: loading the 64-stock Indonesia graph (37,046 triples) into a
standard triplestore (Oxigraph \cite{ref40}) and running the
\textbf{unchanged} SPARQL templates, all five competency questions
return in single-digit milliseconds (Supplementary Table S5). The scaled
cross-sections of Section~\ref{scaling-to-a-64-stock-cross-section} are therefore produced by the system's
own query layer on a real triplestore, byte-identical to the RDFLib
configuration.

\hypertarget{explanation-quality}{%
\subsection{Historical exploratory perception study}\label{explanation-quality}}

Across the three curated markets the inference layer derives 21 findings (9 PSX, 2 MSX, 10 IDX), each with a generated explanation bundle. The final-graph audit separates explicit dependency completeness from structural conformity and source attribution. The original user study evaluates the presentation of evidence-enriched displays rather than validating every provenance link or financial computation.

We conducted an exploratory within-subject perception study. 

\textbf{Seventeen respondents} (6 finance professionals, 5
academics, 3 students, 3 others; self-reported roles) each rated
\textbf{8} OntoKG-EQ statements first as a result-only note (Version A)
and then as the \textbf{same} statement with its provenance-grounded
evidence bundle (Version B), on ad-hoc 7-point \emph{trust} (informed
conceptually by \cite{ref41}, not an adaptation of its scale) and
\emph{justification-completeness} ratings, and chose a preferred
version. Analysed \textbf{at the participant level} (one mean per
condition, n = 17, avoiding pseudoreplication of the 136 item ratings),
mean trust ratings were higher for Version B by +2.87 (2.93 → 5.79;
participant bootstrap 95\% CI {[}2.10, 3.52{]}; exact Wilcoxon
signed-rank p ≈ 7.6×10⁻⁵), as were perceived-completeness ratings by +3.35 (2.10 →
5.44; p ≈ 6.1×10⁻⁵, n = 16 non-zero pairs); \textbf{16 of 17 preferred
the evidence-grounded version on all eight items, and all 17 on a
majority} (sign test p ≈ 1.5×10⁻⁵), with the same direction in the
finance-professional subgroup (six respondents; exploratory trust Δ = +3.42). Table~\ref{tab:t11} reports the overall panel, not a separate subgroup test. These results show higher reported trust and perceived completeness for the richer display in this panel; objective verifiability and calibrated trust were not measured. Because each item presents Version
A before the more-detailed Version B, part of the lift may reflect added
information or order rather than provenance grounding specifically
(Section~\ref{threats-to-validity}). We therefore report this as exploratory evidence of
participant preference under this presentation protocol, not a definitive causal or professional-performance result.

\begin{table}[htbp]
\caption{User study (n = 17; 8 real OntoKG-EQ statements; 7-point scales; within-subject, Version A = result only, Version B = same statement with its evidence bundle). Analysis is at the participant level (one mean per participant per condition) to avoid pseudoreplication.}\label{tab:t11}
\centering\footnotesize\setlength{\tabcolsep}{4pt}
\begin{tabularx}{\linewidth}{XrrXX}
\toprule
Measure & A (mean) & B (mean) & Δ (95\% CI, participant bootstrap) & Test (participant-level, exact) \\
\midrule
Trust & 2.93 & 5.79 & +2.87 [2.10, 3.52] & Wilcoxon p ≈ 7.6×10⁻⁵ (n = 17) \\
Completeness of justification & 2.10 & 5.44 & +3.35 [2.61, 4.00] & Wilcoxon p ≈ 6.1×10⁻⁵ (n = 16) \\
Preferred version — B on all 8 items / on a majority & — & 16/17 · 17/17 & — & sign test p ≈ 1.5×10⁻⁵ \\
\bottomrule
\end{tabularx}
\end{table}

\hypertarget{additional-robustness-checks}{%
\subsection{Additional robustness
checks}\label{additional-robustness-checks}}

The following checks examine structural errors, execution scale, and numerical transformation separately.

\textbf{Fault-injection.} Injecting eight realistic data faults one at a
time into a valid, SHACL-conformant graph (wrong-typed values, missing
names, subjects, and sources, malformed dates, a wrong reporting-period
link, and a duplicate value), \textbf{SHACL detected all 8/8}
(Supplementary Table S6), genuine error-detection evidence rather than a
definitional dependency.

\textbf{Controlled scaling.} To probe scaling beyond the 64-stock graph
we replicated it (distinct instance IRIs) to larger sizes and re-ran the
\emph{unchanged} CQ1/CQ3 templates on the Oxigraph triplestore (Table~\ref{tab:t13}). Ingestion is near-linear (\textasciitilde13 s per million triples).
The single-pattern queries (CQ4, CQ5) stay flat, while the two
multi-join queries grow with the size of their result set, which itself
grows with the graph, yet remain sub-second at \textasciitilde10⁶
triples. These measurements characterize the archived SPARQL query workload
on replicated data, not the new contract checker or source-lineage generation.
The in-memory reference engine remains a limit for this evaluated workload.
(Full-universe throughput, concurrent load, and 10⁷--10⁸ triples on a
clustered store remain future work.)

\begin{table}[htbp]
\caption{Controlled scaling on Oxigraph (replicated 64-stock IDX graph; unchanged CQ1/CQ3 templates; on-disk Oxigraph store, so base-size latencies exceed the in-memory Supplementary Table S5 figures; representative run).}\label{tab:t13}
\centering\footnotesize\setlength{\tabcolsep}{4pt}
\begin{tabular}{rrrr}
\toprule
Triples & Ingest (s) & CQ1 latency (ms) & CQ3 latency (ms) \\
\midrule
37,046 & 0.2 & 9.4 & 11.9 \\
111,130 & 0.8 & 31.8 & 39.4 \\
1,111,264 & 14.4 & 468.6 & 690.5 \\
\bottomrule
\end{tabular}
\end{table}

\textbf{Fidelity audit.} The revised independent check reads actual stored graph literals and compares them with arithmetic sums recomputed directly from the source daily returns. All 64 company window statistics match within $10^{-6}$ percentage points, with maximum absolute error $4.98\times10^{-7}$ percentage points, consistent with six-decimal serialization. Node coverage and company association are also checked. The earlier script compared two source-sheet calculations without reading the graph; the revised audit repairs that verification gap. The check covers these company statistics only, not all financial metrics or original-source authenticity.

\subsection{Archived-graph dependency control}\label{jws-revision-audit}
This subsection and the following legacy contract, corruption and financial requirement-change experiments use archived metric conventions. They are historical controls, not the corrected primary financial evaluation in Section~\ref{strengthened-evaluation}. Their counts must not be combined with corrected-cohort counts.
The revision evaluates the original and repaired final graphs separately, recording input hashes and package versions. All three original final graphs conform to the existing SHACL shapes, yet two PSX R4 findings omit the abnormal-volume observation that caused the rule to select them. After including that observation and regenerating the finding bundles, all 21 findings across PSX, MSX, and IDX pass the additional checks of declared supports, matching windows, rule predicates, and bundle consistency. Final SHACL conformance also holds for all three repaired graphs.

Five controlled interventions preserve the surrounding RDF graph and contextual evidence. The existing shapes detect none of these five changes. The additional dependency audit detects a numeric change that falsifies the selection predicate, a swap to an existing but mismatching analysis window, and removal of one required bundle observation. Neither check detects a plausible numeric perturbation that preserves the predicate or a well-formed wrong source assignment. A known-original snapshot identifies all five changes, but this is an intervention oracle, not independent validation of source truth.

The repaired graphs contain 45 distinct declared supporting observations across the three markets. These are not all necessary operands: an OR condition can have alternative witnesses and attached context. All have temporal context, but the base graphs contain no explicit observation-level source or derivation links for these observations. Their company announcement links provide context, not complete numeric lineage. A separate workbook-lineage overlay is evaluated as a revision artifact; it must not be conflated with original-source authenticity. The dependency checks cover materialized positive findings, not false negatives, every possible CQ answer, or arbitrary derivations.

\subsection{Canonical relational reuse control}\label{jws-relational-control}
The revision implements one constrained SQLite schema, one CQ1 query, one CQ3 query, and one evidence-dependency join without market-specific rewrites. SQL and SPARQL return exactly matching sets, including numeric values, across PSX, MSX, and IDX. CQ1 row counts are 0, 2, and 0; CQ3 counts are 2, 0, and 2; dependency-join counts are 19, 6, and 20 respectively. Numeric CHECK and unknown-entity foreign-key corruption tests are rejected in each database. Schema, query, and input hashes are recorded.

This control imports the same stored graph metrics. It verifies query and join agreement, not independent financial correctness, and empty sets provide weaker evidence than exercised positive cases. Contextual announcement/source joins are retained without treating them as complete numeric provenance. A relational implementation can thus provide constraints, provenance tables, materialized findings, evidence joins and reuse after canonical mapping (Table~\ref{tab:t6}). We make no claim that RDF uniquely enables these functions or reduces engineering effort; OntoKG-EQ expresses them with shared semantic vocabularies, graph identifiers, SPARQL and declarative shapes.

\begin{table}[htbp]
\caption{Capabilities available in both representations. This comparison describes implementation choices, not measured superiority.}\label{tab:t6}
\centering\footnotesize
\begin{tabularx}{\linewidth}{XXX}
\toprule
Dimension & Relational implementation & OntoKG-EQ \\
\midrule
CQ1/CQ3 results & Same results on evaluated inputs & Same results on evaluated inputs \\
Structural constraints & CHECK and foreign-key constraints & SHACL shapes \\
Result evidence & Tables and explicit evidence joins & EvidenceBundle and graph links \\
Derived findings & Rows or materialized views & Typed finding individuals \\
Reuse after canonical mapping & Demonstrated with one schema & Demonstrated with shared semantic artifacts \\
Representation & Relational schema and SQL & RDF vocabulary and SPARQL \\
\bottomrule
\end{tabularx}
\end{table}

\subsection{Workbook-cell lineage overlay}\label{jws-lineage}
A separate PROV overlay links value-matched observations to exact workbook fields identified by file SHA256, worksheet, cell address, row identifiers, and explicit source-register identifiers. Source-register URLs are recorded rather than treated as authenticated sources. Recomputed dependencies follow the actual implemented calculations; numerical agreement is required before a mapping is emitted.

Of 522 PSX observations, 500 have value-matched mappings; the corresponding counts are 319/321 for MSX, 542/542 for IDX, and 4944/4947 for scaled IDX. Every mapped dependency cell has an explicit source identifier resolving to one register row. The 27 unresolved observations across these datasets include values defaulted from missing input cells, and they remain explicitly unresolved. For observations explicitly supporting the 21 repaired findings, 44/45 map to workbook cells and source-register identifiers (PSX 18/19, MSX 6/6, IDX 20/20). The remaining PSX event statistic is zero despite having no supplied event-return cells; its numeric provenance is unresolved. These results establish workbook-level traceability for the mapped observations, not accuracy of original sources or independently verified upstream construction. Correlation mapping uses the existing Pearson helper and is not an independent test of that helper's analytical correctness.

\subsection{Executable contract and bundle replay}\label{contract-replay-results}
The contract checker independently interprets the four declared finding
predicates over typed observations and actual explanation bundles. All 21
inspected findings (9 PSX, 2 MSX, 10 IDX) have sufficient declared support
and a sufficient witness in the correctly targeted bundle. Under this bounded
definition, the complete-bundle rate is 21/21. Each bundle also has at least
one sufficient witness whose recorded dependency cells have consistent source
identifier paths. This does not mean all 45 declared supporting observations
have source mappings: one unresolved zero-valued event statistic is unnecessary
for a passing volume-only R4 witness.

Six synthetic regression tests cover inclusive correlation thresholds,
strict comparison boundaries, missing comparators, same-valued wrong-role
observations, volume-only OR evidence, and bundle-only evidence loss. Decimal
arithmetic preserves lexical decimal boundaries, including a CQ3 company value
of 1.1 against a comparator of 0.1 with a one-point strict margin. Source truth,
absent unit declarations and the unlinked temporal anchors identified in
Section~\ref{executable-contracts} remain unverified. Status and witness triples
are written as separate overlays, with graph, contract, query, rule, shape and
implementation hashes recorded in machine-readable outputs.

\subsection{Controlled semantic corruptions}\label{contract-fault-results}
The expanded suite includes 12 deliberately selected PSX corruptions and
two controls. Every corrupted base graph conforms to the existing SHACL
shapes. Finding-condition replay flags seven corruptions; checking the actual
bundle flags eight; the combined condition, bundle and recorded source-path
policies flag ten. The controls accept the original positive finding and a
volume-only sufficient R4 witness. These counts describe selected interventions,
not a detection-rate estimate for naturally occurring errors.

The cases include another valid company or analysis window, a missing
comparator, bundle-only comparator loss, a wrong metric with its numeric value
unchanged, threshold equality, a result satisfying only one of CQ3's two
comparisons, absent support, a broken source path and an inconsistent source
identifier. A plausible numeric perturbation preserving the predicate remains
undetected, as does a false source attribution whose source edge and identifier
are changed coherently. Independent calculation or trusted source comparison
is required for those defects. Currency/unit corruption is not evaluated because
the observations lack the necessary explicit metadata.

The comparison retains the same RDF evidence while enabling or disabling
these checking policies. It is a policy ablation, not a head-to-head benchmark
against NPCS or another provenance implementation. A separate AND-to-OR query
edit is identified by artifact hash drift; valid stored witnesses alone do not
reveal which query version generated them.

\subsection{Bounded requirement change and replay}\label{contract-evolution-results}
We change CQ3 from exceeding both same-window comparators to exceeding both
by strictly more than one percentage point. Versioned query, documented-rule
and contract artifacts preserve the old and new predicates. The new SELECT
and CONSTRUCT results agree with an independent imperative traversal of the
same observations: PSX changes from two selected companies to one, IDX from
two to one, and MSX remains empty. All four archived R3 finding supports replay
consistently with the corresponding old/new selection conditions.

Five synthetic boundary cases yield three original selections and one revised
selection; equality at either strict threshold is excluded. These fixtures are
not market observations. Replaying the archived original query reproduces its
results, while pairing it with the changed contract fails the expected query
hash. CQ1, CQ2 and CQ4 are read-only same-snapshot controls with unchanged
results; they do not establish general regression freedom. Artifact impact is
declared for this threshold edit using manifest paths, rather than inferred by
a general change-impact algorithm. The production inference code and market
graphs remain unchanged; adopting the revised threshold would require deploying
the corresponding inference version.

\hypertarget{discussion-and-threats-to-validity}{%
\section{Discussion and threats to
validity}\label{discussion-and-threats-to-validity}}

\hypertarget{interpretation}{%
\subsection{Interpretation}\label{interpretation}}

The central result is a bounded executable connection between analytical
conditions and sufficient typed evidence in actual finding bundles. The NHS
application shows that this connection can be applied to an externally
specified public statistical task with source reconciliation and explicit
missingness. The semantic corruption suite demonstrates distinctions that the
existing shapes do not encode, while the financial margin and NHS policy
replays demonstrate reproducible evaluation of declared requirement changes.
These are task-specific checks,
not new general provenance or ontology-reasoning algorithms. Reuse of shared
artifacts after canonical mapping remains useful engineering evidence without
establishing automatic interoperability or reduced onboarding effort relative
to a canonical relational schema.

The intended benefit is inspectability with explicit failure states: insufficient
roles, an incomplete actual bundle, and inconsistent recorded source paths can
be reported separately. The 21/21 sufficient-witness result covers inspected
positive findings under declared predicates, not false-negative discovery,
original-source authentication, or full verification of every temporal and
unit assumption. Matching result counts across curated slices do not establish
generality, and perceived trust should not be interpreted as calibrated trust
or improved professional audit performance.

\hypertarget{threats-to-validity}{%
\subsection{Threats to validity}\label{threats-to-validity}}

\textbf{Construct validity.} The derived metrics are standard analyst
proxies. The sector peer-baskets and broad-market benchmarks are
transparent demonstrator constructions, documented per market, not
licensed index feeds. The CQ4/CQ2 thresholds (\textbar CAR\textbar{} ≥
2\%, volume ratio ≥ 1.5, \textbar correlation\textbar{} ≥ 0.3) are fixed
a priori as an operational screen informed by event-study practice
\cite{ref35,ref36,ref37} and are not tuned to the data. Sector comparators are
equal-weighted demonstrator baskets, event dates anchor on the reported
announcement date without special after-hours handling, and closes are
unadjusted for corporate actions. These are documented demonstrator
simplifications, to be replaced with licensed, adjusted feeds before any
empirical market claim.

\textbf{Internal validity.} Deterministic execution and SHACL gating do not
establish independent source accuracy. The boundary benchmark is constructed
to isolate known failure modes; it does not estimate their natural prevalence.
The first matched comparator covers the synthetic sensor OR configuration;
an additional analytical suite covers all four configured finding families,
numerical status and unit readiness under declared diagnostic parameters.
Neither establishes equivalence for arbitrary RDF literals, expressions or
full SHACL. The large financial declarative replay uses explicit in-memory
adapters and direct pool evaluation, while the smaller release comparison
exercises the audit entry point after interface adaptation. Neither is an
unadapted replacement of the archived pipeline. Small slices and overlapping
cohort versions must not be treated as independent empirical samples.
The NHS application independently reconstructs a genuine publisher-defined
task, but its CSV and workbook originate from the same reporting collection.
Agreement verifies reconciliation with the inspected published aggregates,
not underlying patient records, clinical quality or all upstream measurements.
Its explanation views are generated by this application; no naturally occurring
NHS explanation omission or practitioner performance gain is established.

\textbf{External validity.} The curated per-market work uses bounded
slices (two sectors, two firms each); behaviour on large constituent
universes is exercised only through the 64-stock Indonesia
cross-section. The multi-join CQ latency that is engine-bound on the
in-memory reference store is resolved on a standard triplestore (Section~\ref{scalability}, Supplementary Table S5), so the open question is ontological
fidelity at full-exchange breadth, not execution. Three independent
markets and real, exactly-intersected Indonesian data mitigate breadth,
but generalization to full exchanges remains future work. The separate NHS
application provides real nonfinancial aggregate-data evidence for the bounded
audit interface, not for reuse of the financial ontology or inference rules.
One month, one indicator and an author-implemented adapter do not establish
independent adoption, longitudinal reliability or general domain portability.

\textbf{Conclusion validity.} Structural conformity, dependency completeness, and source fidelity are distinct claims. The within-subject user study (n = 17, Section~\ref{explanation-quality}) provides exploratory evidence of higher perceived trust and completeness for evidence-enriched displays (both p \textless{} 0.001), rather than establishing objective audit performance or calibrated trust.
Because each item presented the result-only version before the
evidence-grounded one, part of this lift may be an order effect; the prepared
counterbalanced objective study (Section~\ref{objective-study}) is designed to
address this.

\textbf{Scope.} OntoKG-EQ is a data-and-knowledge-engineering
contribution. It makes no predictive or causal-economic claim;
``outperformance'' and ``divergence'' are defined operationally over the
stated windows and metrics, not as economic causation.

\hypertarget{design-implications}{%
\subsection{Design implications}\label{design-implications}}

The contribution is a specialized, reproducible audit system and diagnostic
protocol, rather than a new sufficiency logic or discovery of graph-scope
effects. Matched functional agreement establishes that the implemented
interface respects the tested conditions; it does not establish lower
authoring effort, superior usability or greater accuracy than an equivalently
specified validator. Those practical advantages require independent deployment
or objective user evidence beyond the current evaluation. The value of the
artifacts is the explicit, inspectable pairing of finding, support and delivery
outcomes under analytical conditions, with these limitations retained.

A further distinction is between sufficient evidence in one selected snapshot
and support across declared source versions. A separate release-robustness
extension keeps values from the same publication together and asks whether a
fixed comparison remains positive in every admitted release. Two historical
BEA examples illustrate a comparison that reverses between releases and another
whose positivity is obscured by independent marginal bounds
\cite{beaQ1Second,beaQ1Third,beaQ2Sep,beaQ2Oct}. Online Resource~1, section
``Release-coherent robustness of published comparisons,'' gives the source
bindings, restricted procedure and replay. This is not an alteration of the
three-valued boundary audit or source authentication. Existing provenance
systems support dependent aggregate comparisons~\cite{provsqlAggregation};
the extension is a reproducible specialization, not a claim to introduce
possible-world semantics or a new general uncertainty method.

The evidence bundle can support deterministic presentation of values and sources, with optional natural-language assistance. However, even connective prose may introduce unsupported causal, comparative, or temporal assertions. A hybrid renderer would require separate claim-level checking and evaluation; the current results do not establish that such a design achieves perfect faithfulness.

Publishing CQ definitions, artifact mappings, constraints and result-evidence links lets others inspect the intended contract; demonstrated reuse remains conditional on canonical mapping.

\hypertarget{limitations-and-future-work}{%
\section{Limitations and future
work}\label{limitations-and-future-work}}

The most important limitation is \textbf{evidential scope}. The financial
data are fixtures with demonstrator comparators and partially reviewed sources,
so they test stored-data semantics rather than authenticated market effects.
Licensed, adjusted index and constituent feeds and complete primary-source
review are required before any empirical market claim. The NHS application
covers one month, one indicator and an author-implemented adapter; repeated
releases, further publishers and independent developers are needed to assess
longitudinal reliability and onboarding effort.

\textbf{Ontological fidelity at breadth} remains untested. The three markets
share essentially the same data shape, so 100\% apparatus reuse shows that one
ontology fits three \emph{structurally similar} markets, not genuinely
heterogeneous ones. Mapping raw sources into the nine-sheet schema is manual and
unmeasured; onboarding by an independent developer is the decisive portability
test.

\textbf{Human benefit} is unestablished. The historical ratings measure
preference only. The prepared counterbalanced objective study
(Section~\ref{objective-study}) must be run, after ethics review, to test
decision accuracy, source identification and completion time. A model
transcription experiment would need archived answers and claim-level scoring.
The frozen competency-question set can grow under the same governance, each new
term justified by a new question.

\hypertarget{conclusion}{%
\section{Conclusion}\label{conclusion}}

OntoKG-EQ organizes analytical knowledge-graph construction around explicit
competency questions and audits finding evidence through executable contracts
that check ambient observations, declared support and delivered observations
separately. On a public NHS statistical screen, independent source
recomputation reconciles 532 providers with the publisher's aggregates, keeps
one roster entry unavailable, and agrees with a matched SHACL-SPARQL baseline on
all 533 frames; six replayed threshold policies identify their exact affected
providers without core-interpreter changes. Constructed diagnostics yield 78
status/readiness agreements over 26 analytical cases, actual release replay
reproduces two support/delivery omissions (19/21 to 21/21 after repair), and
487 paired financial records agree under disclosed adapters. In the corrected
64-company financial cohort, all 16 complete CQ1/CQ3 comparisons match
independent traversal, but only 22 of 42 numerical positives satisfy the
metadata policy, and source audits expose scaling and temporal defects.

These results support bounded evidence sufficiency, explicit failure reporting
and reproducible artifact versions, built from established RDF, SPARQL, SHACL
and provenance mechanisms. They do not establish source truth, an advantage over
an equivalently specified validator or relational implementation, authenticated
market effects, independent deployment or objective human audit benefit. A
predicate-preserving numeric error and a coherently false source assignment
remain outside the contract checks, and the prepared human experiment has not
been executed.

\hypertarget{statements-and-declarations}{%
\section{Statements and
declarations}\label{statements-and-declarations}}

\textbf{Funding.} This research received no specific grant from any
funding agency in the public, commercial, or not-for-profit sectors.

\textbf{Competing interests.} The authors have no competing interests to
declare that are relevant to the content of this article.

\textbf{Ethics and consent.} The historical perception study
(Section~\ref{explanation-quality}) was an anonymous online questionnaire
completed by seventeen adult volunteers between 7 and 19 July 2026. Each
participant confirmed being 18 or older and gave consent in the form's first,
mandatory item; no names or contact details were collected, and results are
reported only in aggregate. No ethics review was sought before
data collection. On 1 November 2026, after data collection, the
Research-Ethics Coordinator of the Department of Computer Science, City
University of Science and Information Technology, confirmed in writing
(ref.\ CUP-26-15E) that under institutional policy the study does not require
ethics committee review and that its aggregate results may be published.
Participant-level
records are kept private. The prepared objective study has not recruited
participants. The NHS application uses public provider-level aggregates, not
individual patient records.

\textbf{Data and code availability.} The original reproducibility
package (ontology, SHACL shapes, the five competency-question SPARQL
templates, inference rules, the unified builder and
inference/explanation generator, per-market data and materialized
graphs, evaluation scripts and outputs, a data dictionary, and the
recorded environment) is openly available on GitHub
(\url{https://github.com/furqan-nr/OntoKG}) and archived on Zenodo
(\url{https://doi.org/10.5281/zenodo.21569316}), including the Indonesia data-acquisition scripts and exact-date intersection builder. The revision code, contracts, corrected outputs, lineage overlays and NHS application are released in the same repository as v2.0.0 (\url{https://github.com/furqan-nr/OntoKG/releases/tag/v2.0.0}) and archived on Zenodo (\url{https://doi.org/10.5281/zenodo.23860671}). Public NHS originals are linked by exact URL and hash in the source manifest; attribution follows the Open Government Licence v3.0, with logo and third-party exceptions. Captured third-party source documents and inputs whose redistribution rights are not established are not redistributed; they are identified by URL and hash. Participant-level survey records are not publicly shared because consent specifies aggregate reporting.

\textbf{Author contributions (CRediT).} \textbf{Furqan Nasir:} Conceptualization;
Methodology; Software; Validation; Formal analysis; Data curation;
Writing -- original draft; Visualization. \textbf{Muhammad Atif Saeed:}
Methodology; Writing -- review \& editing. \textbf{Muhammad Ehsan:}
Methodology; Writing -- review \& editing. \textbf{Sher Jeel Ahmad:}
Validation; Writing -- review \& editing. \textbf{Abdul Moiz Altaf:}
Data curation; Writing -- review \& editing.

\vspace{4pt}\noindent\textbf{Supplementary information.} Online Resource 1 distinguishes historical supporting tables from the revision checks. Revision scripts and machine-readable reports are part of the released archive.

\hypertarget{declaration-of-generative-ai}{%
\section{Declaration of generative AI and AI-assisted technologies in the manuscript preparation process}\label{declaration-of-generative-ai}}

The original manuscript records use of Anthropic Claude during 2025--2026
for prose and selected scripts. The revision additionally used OpenAI
Codex assistants in October 2026 to review positioning, develop and execute
evaluation scripts, inspect outputs, repair evidence-support generation,
and revise the manuscript. New reported quantities are computed from the
documented inputs by the accompanying scripts. All authors reviewed and
approved the revised manuscript and take full responsibility for its
content; AI assistance does not establish source accuracy.

\clearpage
\appendix
\section*{Online Resource 1: Supplementary material}\label{online-resource-1}
\lstset{basicstyle=\scriptsize\ttfamily}
\noindent This appendix is the supplementary material (Online Resource 1) referred to in the main text.

\renewcommand{\thesection}{\Alph{section}}\setcounter{section}{0}
\hypertarget{llm-prompt-and-decoding-settings}{%
\section{Historical model-transcription experiment (excluded)}\label{llm-prompt-and-decoding-settings}}

This section documents the historical transcription experiment, which is excluded from the revised main quantitative evaluation. Eight instruction-tuned checkpoints received a structured evidence context and generated brief answers using greedy decoding. The notebooks and aggregate scores do not archive the generated answer texts needed for retrospective rescoring. No new model scores are inferred from those aggregates.

A future reproducible run must save the exact context and answer, graph hash, case identifier, source identity, prompt, model revision, and scorer version. Contextual company announcements must be distinguished from the actual sources of numeric observations.

\hypertarget{competency-question-sparql-templates-cq1-and-cq3}{%
\section{Competency-question SPARQL templates (CQ1 and
CQ3)}\label{competency-question-sparql-templates-cq1-and-cq3}}

The two multi-join analytical queries are inlined below exactly as
executed, unchanged, on both the RDFLib reference engine and the
Oxigraph triplestore (main paper, Scalability subsection). The full CQ1--CQ5 set, the SHACL
shapes, and the R1--R4 CONSTRUCT rules are released verbatim in the
reproducibility package (\passthrough{\lstinline!queries/!},
\passthrough{\lstinline!shacl/shapes.ttl!},
\passthrough{\lstinline!rules/!}).

\begin{lstlisting}[language=SPARQL]
# CQ1 — stronger reported fundamentals, weaker market response than the benchmark
PREFIX : <https://w3id.org/ontokg-eq#>
SELECT ?company ?yoyProfitGrowth ?companyReturn ?benchmarkReturn WHERE {
  ?fg :isObservationOf ?company ; :hasMetricName "YoY profit growth %" ; :hasMetricValue ?yoyProfitGrowth .
  ?company a :Company .
  ?cr :isObservationOf ?company ; :hasMetricName "post-report window return %" ;
      :hasMetricValue ?companyReturn ; :observedOverWindow ?w .
  ?br :isObservationOf ?bm ; :hasMetricName "benchmark window return %" ;
      :hasMetricValue ?benchmarkReturn ; :observedOverWindow ?w .
  ?bm a :MarketIndex .
  FILTER(?yoyProfitGrowth > 0 && ?companyReturn < ?benchmarkReturn)
} ORDER BY DESC(?yoyProfitGrowth)
\end{lstlisting}

\begin{lstlisting}[language=SPARQL]
# CQ3 — outperforms BOTH sector comparator AND broad-market benchmark over the same window
PREFIX : <https://w3id.org/ontokg-eq#>
SELECT ?company ?companyReturn ?sectorReturn ?benchmarkReturn WHERE {
  ?cr :isObservationOf ?company ; :hasMetricName "post-report window return %" ;
      :hasMetricValue ?companyReturn ; :observedOverWindow ?w .
  ?company a :Company ; :isClassifiedByIndustrySector ?sector .
  ?sr :isObservationOf ?sector ; :hasMetricName "sector window return %" ;
      :hasMetricValue ?sectorReturn ; :observedOverWindow ?w .
  ?br :isObservationOf ?bm ; :hasMetricName "benchmark window return %" ;
      :hasMetricValue ?benchmarkReturn ; :observedOverWindow ?w .
  ?bm a :MarketIndex .
  FILTER(?companyReturn > ?sectorReturn && ?companyReturn > ?benchmarkReturn)
} ORDER BY DESC(?companyReturn)
\end{lstlisting}

\hypertarget{automatic-scorer-specification-and-validation}{%
\section{Historical scorer and its limitations}\label{automatic-scorer-specification-and-validation}}

The historical scorer measured numeric precision, lexical unsupported-assertion cues, and a provenance indicator. The currently released code adds numeric-value recall and accepts a matching date or source token. These definitions must not be conflated with the historical aggregate scores.

Numeric precision can assign a perfect score to an answer with no numbers. Distinct-value recall does not distinguish two different metrics with the same numeric value. A matching date does not establish correct source identity, and lexical cue detection does not establish entailment. Since original answer texts are not archived, stronger semantic scoring cannot be reconstructed for the historical panel. The revision report \path{evaluation_jws/llm_rescore.md} records this limitation.

\hypertarget{extended-formal-model}{%
\section{Extended formal model}\label{extended-formal-model}}

This section gives the full derived-metric and inference-rule detail
summarised in Section 3.2.

\textbf{Derived metric.} A derived metric is a function
\(\varphi : (2^{\mathrm{Obs}}, W) \to \mathrm{Obs}\) that maps a set of
base observations and a window \(W\) to a new (derived) observation.
OntoKG-EQ defines nine CQ-justified derived metrics: post-report window
return, year-on-year profit growth, sector window return, benchmark
window return, exchange-rate window change, exchange-rate-versus-company
return correlation, exchange-rate-versus-benchmark return correlation,
the event-window sum of supplied returns (historically labelled cumulative abnormal return), and abnormal volume ratio. Derived observations are first-class observations; their declared structural constraints, deciding dependencies, and source-cell lineage require separate checks.

\textbf{Inference rule.} An inference rule
\(\rho : \mathrm{Pattern} \to \mathrm{Finding}\) maps a graph pattern
encoding a CQ condition over derived metrics to an AnalyticalFinding
\(f = (e, t, \rho_{\mathrm{id}}, q, O_f)\), where \(e\) is the result
entity, \(t\) the finding type, \(\rho_{\mathrm{id}}\) the rule
identifier, \(q\) the CQ family, and \(O_f\) the set of observations
whose values triggered the rule.

\begin{center}\rule{0.5\linewidth}{0.5pt}\end{center}

\setcounter{table}{0}\renewcommand{\thetable}{S\arabic{table}}
\section{Supplementary tables}
\begin{table}[htbp]
\caption{Per-market adapter configuration: the analytical keys that differ between markets (sheet-file prefix, data folder and display labels are also configured). The ontology, shapes, queries, rules and code are unchanged.}\label{tab:tS1}
\centering\footnotesize\setlength{\tabcolsep}{4pt}
\begin{tabularx}{\linewidth}{XXXX}
\toprule
Adapter key & PSX & MSX & IDX \\
\midrule
namespace & .../psx\# & .../msx\# & .../idx\# \\
company identifier column & company\_\allowbreak symbol & stock\_\allowbreak code & stock\_\allowbreak code \\
EPS column & eps & eps\_\allowbreak myr & eps \\
FX rate column / sheet & pkr\_\allowbreak usd\_\allowbreak rate / 06\_\allowbreak sbp\_\allowbreak fx & myr\_\allowbreak usd\_\allowbreak rate / 06\_\allowbreak bnm\_\allowbreak fx & idr\_\allowbreak usd\_\allowbreak rate / 06\_\allowbreak fx \\
local currency & PKR & MYR & IDR \\
growth-units scale & ×1 (percent) & ×100 (ratio) & ×1 (percent) \\
benchmark comparator type & market index & broad\_\allowbreak market\_\allowbreak benchmark\_\allowbreak proxy & broad\_\allowbreak market\_\allowbreak benchmark \\
benchmark label & KSE-100 & FBM KLCI (proxy) & Jakarta Composite (JCI) \\
\bottomrule
\end{tabularx}
\end{table}

\begin{table}[htbp]
\caption{(PSX, MSX, and IDX are our internal dataset codes for the Pakistan Stock Exchange, Bursa Malaysia, and Indonesia Stock Exchange datasets, respectively.) Markets and constituents.}\label{tab:tS2}
\centering\footnotesize\setlength{\tabcolsep}{4pt}
\begin{tabularx}{\linewidth}{lXXX}
\toprule
Market & Exchange / benchmark & Sector A (2 firms) & Sector B (2 firms) \\
\midrule
PSX & Pakistan Stock Exchange / KSE-100 & Fertilizer: ENGROH, FFC & Oil \& Gas Exploration: OGDC, PPL \\
MSX & Bursa Malaysia / FBM KLCI (proxy) & Financial Services: MAYBANK, CIMB & Telecommunications \& Media: TM, MAXIS \\
IDX & Indonesia Stock Exchange / Jakarta Composite & Banking: BBCA, BMRI & Telecom: ISAT, TOWR \\
\bottomrule
\end{tabularx}
\end{table}

\begin{table}[htbp]
\caption{Historical timing comparison omitted from the main evaluation.}\label{tab:tS3}
\centering
\begin{tabularx}{\linewidth}{XX}
\toprule
Measurement & Interpretation \\
\midrule
Template retrieval/rendering and model generation on different hardware & Different tasks; no like-for-like efficiency advantage established \\
\bottomrule
\end{tabularx}
\end{table}

\begin{table}[htbp]
\caption{Component ablations on the PSX and MSX graphs (each component removed in isolation).}\label{tab:tS4}
\centering\footnotesize\setlength{\tabcolsep}{4pt}
\begin{tabularx}{\linewidth}{XXrr}
\toprule
Removed component & Metric affected & Full & Ablated \\
\midrule
Provenance layer & historical evidence availability & 1.00 & 0.00 \\
Evidence-source links & evidence coverage & 1.00 & 0.00 \\
Inference rules & derived findings (PSX / MSX) & 9 / 2 & 0 / 0 \\
Evidence bundles & CQ5 explainability rows (each market) & 3 & 0 \\
SHACL & malformed observation detected & yes & no \\
\bottomrule
\end{tabularx}
\end{table}

\begin{table}[htbp]
\caption{The unchanged CQ1–CQ5 SPARQL templates on the 64-stock Indonesia graph (37,046 triples), executed on the Oxigraph triplestore. Latencies are medians of 20 runs (min–max in brackets) after a 70 ms one-time load, for a representative run, and are hardware- and run-dependent. The first-call JIT warm-up inflates the maximum. CQ1 and CQ3 are the multi-join queries that do not complete on RDFLib's in-memory engine at this size.}\label{tab:tS5}
\centering\footnotesize\setlength{\tabcolsep}{4pt}
\begin{tabularx}{\linewidth}{XrX}
\toprule
Query & Rows & Latency (ms): median [min–max] \\
\midrule
CQ1 (fundamentals vs market response) & 17 & 3.8 [3.7, 4.3] \\
CQ2 (FX association) & 63 & 0.3 [0.2, 0.6] \\
CQ3 (outperform sector and benchmark) & 24 & 5.0 [4.9, 5.6] \\
CQ4 (announcement event windows) & 0 & 1.6 [1.6, 2.9] \\
CQ5 (explainability / provenance) & 3 & 0.1 [0.1, 0.1] \\
\bottomrule
\end{tabularx}
\end{table}

\begin{table}[htbp]
\caption{Fault-injection: injected data errors and whether SHACL detects them (valid PSX graph, one fault per trial).}\label{tab:tS6}
\centering\footnotesize\setlength{\tabcolsep}{4pt}
\begin{tabularx}{\linewidth}{XX}
\toprule
Injected fault & Detected by SHACL \\
\midrule
wrong metric-value datatype (decimal→string) & yes \\
company missing name & yes \\
market observation missing subject link & yes \\
malformed announcement date (date→string) & yes \\
observation missing metric name & yes \\
fundamental linked to wrong reporting-period class & yes \\
duplicate metric value (violates maxCount) & yes \\
evidence item missing source & yes \\
\bottomrule
\end{tabularx}
\end{table}
\section{Final-graph, relational and lineage checks}
The revised inference code includes abnormal-volume observations in R4 support and validates the final graph after generating findings and bundles. All 21 findings across the three curated markets pass the added dependency checks. The original PSX graph contained two volume-triggered findings lacking that support.

Five final-graph mutations retain contextual evidence. Existing SHACL detects 0/5, while the additional dependency audit detects 3/5. A source swap and a plausible numeric change preserving the predicate remain undetected. Known-original snapshot checks are intervention oracles, not independent source verification.

The independent source-to-graph return audit verifies 64/64 arithmetic company window statistics within $10^{-6}$ percentage points. It does not verify compounded holding-period returns, all metrics, or source truth. The event builder sums supplied event returns without benchmark subtraction. For IDX, the acquisition assembler writes ordinary consecutive-close company returns to these rows, so the archived event statistic is not market-adjusted CAR; PSX/MSX upstream interpretation remains unverified.

One unchanged constrained SQLite schema and query set matches SPARQL CQ1/CQ3 and evidence-dependency joins on PSX, MSX, and IDX. This establishes canonical relational reuse as a capable control, not RDF superiority.

The separate PROV overlay maps 44/45 deciding observations to exact workbook cells and registered source identifiers. One missing-input PSX event statistic remains unresolved. Full observation mapping counts are 500/522, 319/321, 542/542, and 4944/4947 for PSX, MSX, IDX, and scaled IDX. Workbook lineage does not authenticate original documents or numeric data.

Exact graph hashes, source-cell mappings, script versions, and detailed checks are recorded in the machine-readable reports under \path{reproducible_OntoKG/evaluation_jws/}. They are part of the released revision archive.

\section{Executable evidence contracts and semantic interventions}
The machine-readable registry specifies R1--R4 metric roles, entity roles,
compatible windows, predicate parameters, expected units and related artifact
paths. R2 concerns the FX-sensitive classification rather than the full CQ2
answer. CQ5 supplies the explanation policy. The checker uses Decimal values
and independently replays declared finding support and observations available
in the correctly targeted actual bundle. R4 admits either passing OR branch;
unused operands are context rather than necessary evidence. Missing graph units
and undeclared temporal anchors remain unverified.

All 21 inspected bundles pass the bounded sufficient-witness policy, and each
has a witness whose recorded cell-to-source paths carry matching identifiers.
This does not authenticate sources or prove that all upstream computational
dependencies have been recorded. Six synthetic tests cover threshold equality,
role identity, missing evidence, actual-bundle integrity and OR sufficiency.

\begin{table}[htbp]
\caption{Selected semantic corruption suite: 12 corruptions plus two acceptance controls. Counts are not estimates of naturally occurring error detection.}
\centering
\begin{tabularx}{\linewidth}{Xr}
\toprule
Policy & Corruptions flagged (of 12) \\
\midrule
Existing SHACL shapes & 0 \\
Finding-condition replay & 7 \\
Actual-bundle sufficiency & 8 \\
Combined condition, bundle and source-path policies & 10 \\
\bottomrule
\end{tabularx}
\end{table}

The two undetected corruptions preserve the predicate despite a small numeric
change, or alter both a source edge and identifier to create an internally
consistent false attribution. A separate inconsistent source edge is detected.
The same-data control disables contract policies; it is not an evaluation of
NPCS or another external provenance implementation. Query AND-to-OR hash drift
is reported separately from witness validation. Currency/unit corruption is
not tested because explicit observation metadata is absent.

\section{Financial requirement-change scenario}
Changing CQ3 to require more than one percentage point over both comparators
reduces selections from two to one in PSX and IDX; MSX remains empty. Both
versioned SELECT and CONSTRUCT results match independent graph traversal.
Five synthetic boundary fixtures select three companies under the original
predicate and one under the revised predicate. Old/new contracts, queries,
rules and hashes are preserved under the contract-evolution outputs.
Read-only unchanged-CQ controls do not establish a general regression theorem.
Adopting the revised threshold in production would require deploying the
corresponding inference implementation.

The complete local verification workflow is provided by
\path{reproducible_OntoKG/run_jws_revision.py}. It regenerates audit
artifacts without rebuilding or replacing the archived market graphs.
\section{Strengthened revision artifacts}
The financial-v2 graphs are separate from archived inputs. See \path{evaluation_jws/financial_v2/} for formulas, missing-data policies, graph hashes, and primary-source spot checks; \path{expanded_evaluation/} and its categorical rerun \path{expanded_evaluation_v2/} for all cases, complete-query oracle comparisons and evidence-withholding variants (330 in total across paired cohorts); and \path{metadata_v2/} for declaration checks and failures. The corrected scaled cohort has 42 numerical positives but only 22 metadata-qualified positives. Its 147 negatives, three incomplete classifications and 64 event-not-applicable cases remain separately visible. Corrected curated positives are 16 numerically and 15 metadata-qualified; overlapping cohorts are not independent replications.

\path{external_provenance/} retains original and generated queries, software identities and executed NPCS results. The unmodified native FILTER path fails on the tested CQ3 queries; reported successful comparisons concern BGP provenance with disclosed post-filtering. This is not a replacement of the earlier controlled policy ablation by an external fault-detection experiment.

\path{objective_study_v2/} contains the prepared, unregistered protocol, four counterbalanced packets, facilitator answer key (withheld from the public release) and empty response template. Twelve distinct tasks contain identical source material across displays. No objective participant results exist. The analysis rejects the empty template; synthetic software controls are never counted as participants. Responses and participant-level analysis must be kept private according to consent and ethics requirements.
\section{Current source-assurance artifacts}
The current annual-fact evidence combines six OGDC repairs in \path{annual_source_repair/} and six completed PPL repairs in \path{ppl_source_completion/}. The twelve facts include eight monetary scale corrections and four direct EPS agreements. Earlier source-period failures and pending states remain historical records, not current unresolved PPL statuses. Ten PPL completion checks and five annual-repair mutation checks pass. These reruns validate preserved source bytes and reviewed transcriptions; they do not independently interpret new documents.

The \path{temporal_resolution/} overlay quarantines twenty scaled R1 recent-analysis cases and the curated TOWR mismatch. FFC's issuer document date and Maybank's annual-report announcement match their anchors, but no first-earnings publication timestamp is authenticated. Exact source URLs, scopes and date definitions are in the acquisition history below.

The \path{authentic_prices_malaysia/} capture supplies 32 required Maybank company dates and matches all thirty original post-report closes. The issuer republication's adjustment convention remains unresolved. The separate \path{authentic_prices_psx/regulator_recovery.json} verifies eight unique observations from four original SECP reports: six KSE100 total-return levels and two FFC closes. The earlier indexed-only ledger is historical. Nine price-recovery controls pass. Missing FFC baselines, missing KSE100PR/KLCI series and corporate-action definitions prevent valid market-relative classification. Captured PDFs, renderings and CSV remain local pending redistribution-rights review; hashes are integrity checks, not publisher signatures.

\section{Frozen source-review frame and consolidated cohort}
The \path{source_sampling/} manifest fixes 52 cases across 29 corrected dataset/rule/status strata using seed 20261016. The exact source report is preserved as \path{expanded_evaluation_frozen.json} with a matching SHA256. This frame includes positive, negative and incomplete classifications; it was frozen after purposive checks and is not preregistration. Selected cases are not assumed authenticated. Shared dependencies, unequal sampling fractions and overlapping IDX cohorts prohibit an unweighted pooled source-accuracy estimate.

The separate \path{source_backed_cohort/} replay consolidates FFC and Maybank as subsets of the eight-candidate roster. Six controls check endpoint arithmetic, missing/conflicting observations, horizons and refusal to promote absent benchmark results. Both source-qualified labels remain unknown, not negative, and the two cases do not increase the independent sample. A feed-specific adjustment legend cannot be transferred to another provider. Details of the definition audit are in \path{price_definition_audit/}.

The executed \path{systematic_source_audit/} reports 2,357 unique recorded workbook-cell inputs across the frozen 52 cases. Coverage counts are 2,271 unreviewed lineage inputs, 46 declared proxies, thirty issuer-republished unknown-basis closes, four exact scale-corrected PAT facts, two rounded narrative agreements, two regulator numeric agreements with unresolved workbook basis, and two price disagreements. The thirteen assertions pass; all four input graph hashes match the frozen frame. No case has complete coverage under the all-recorded-input criterion. This does not test minimal source-authenticated AND/OR sufficiency or prove that an unreviewed case is false. All reviewed evidence remains purposive; no pooled population error-rate estimate is made.

\section{Public NHS application artifacts}
The separately versioned \path{external_applications/nhs_rtt_v1/}
contains the frozen task protocol, source manifest, independent source-oracle
records, adapter records and application outputs. The source archive and
provider workbook are locally retained originals, protected by SHA256 hashes;
they are not silently overwritten or included in a public endorsed NHS release.
The manifest supplies exact publisher URLs and OGL attribution with logo and
third-party exceptions. CSV and workbook are different representations of the
same reporting collection, not independent collections.

The source oracle and source adapter separately parse the original CSV and
workbook without importing the RDF checker. They reconcile all 532 eligible
provider records; the extra workbook-only \texttt{D1V5N} roster entry is
explicitly unavailable from selected CSV rows, not zero-filled. The adapter
generates two deciding count roles and recorded row/column selectors, while
the contract compiles fixed factors 100 and 92 into the existing addition
expression language. The interpreter itself is unchanged. The dated
implementation addendum explains why the protocol's anticipated optional
arithmetic extension proved unnecessary; the frozen protocol remains intact.

The actual SHACL-SPARQL/availability comparator reconstructs its pools from RDF
without checker selection helpers. It agrees on all 533 frames, including
the unavailable record; 533 actual SHACL validations are recorded, with
identical within-frame pools cached. Three matching pool labels are not three
independent samples. Source cells, generated triples and code/configuration
hashes document execution and engineering scope, not human time savings.
The evidence views are generated by this application and contain no patient
records. No natural reconciliation defects or objective human outcomes are
claimed. 

\section{Public-data requirement-change replay}
The protocol is \path{contracts/nhs_change_protocol_2026-10-19.json}.
Modules \path{docs/jws_nhs_change_oracle.py},
\path{docs/jws_nhs_change_baseline.py} and
\path{docs/jws_nhs_change_replay.py} separate integer expectations, independent
parameterized queries and interpreter replay. Four regression methods in
\path{docs/test_jws_nhs_change_replay.py} pass. Outputs in
\path{evaluation_jws/nhs_change_replay_v1/} retain each contract, exact queries,
all provider identities, affected sets, source-derived artifact/code hashes
and the execution-plan note. Six policies reuse one cohort; 3,198
provider-policy evaluations are not 3,198 independent observations. Fourteen
constructed controls per policy yield 252 actual SHACL pool validations,
separately from provider-bound cohort SPARQL. The zero-denominator software
guard returns false for the guarded predicate, not a defined percentage or a
failed target for an unavailable provider.

Both implementations parameterize the rule and agree with the integer oracle.
Original-policy re-execution restores outputs without editing the interpreter
or earlier source-derived artifacts. This is not stateful deployment rollback,
a human maintenance comparison or a new NHS policy. The interrupted initial
query attempt supplies no timing claim. The protocol predates variant execution,
not external preregistration or blinding. A parent-side independent integer
calculation reproduces every provider status and exact affected set; a separate
review verifies the original ZIP/XLSX sizes and hashes at review time. These
are implementation/source-identity checks, not new data collection or external
practitioner validation. Unit readiness concerns declared schema consistency,
not independent verification of clinical measurement accuracy.

\section{Release-coherent robustness of published comparisons}
\label{supp:release-robustness}
\paragraph{Task and interpretation.}
Evidence sufficiency in a selected snapshot does not establish stability across
source versions. A separate prototype asks whether a fixed, locally qualified
comparison passes in every admitted whole-release alternative. Its semantics is
$\forall\omega\in\Omega\;\exists W_\omega$ such that $W_\omega$ is a passing
qualified witness; witnesses may change between worlds. The outputs are
\texttt{certain-qualified-positive} or an explicit non-positive counterworld.
Missing roles can cause non-positivity, so a counterworld is not automatically
the main interpreter's numerical \emph{false}. This prototype neither modifies
that interpreter nor independently authenticates source measurements.

Two purposively selected BEA examples use published one-decimal real growth
rates for the United States, percent change from the preceding quarter at
seasonally adjusted annual rates. Each case has one source group: a mode
contains both values from one dated release. Superseded estimates are
historical versions, not equally valid current truths. The analyst-defined
question is retrospective robustness over the listed initial releases, not
an official policy test, causal inference or robustness to all later updates.
Published decimals are exact inputs here; unpublished digits and rounding
uncertainty are not modeled.

\begin{table}[htbp]
\centering\small
\caption{Selected historical release comparisons. Each pair is centre/comparator
growth in percent; rows are coherent release modes, not independent samples.}
\label{tab:release-robustness}
\setlength{\tabcolsep}{3pt}\scriptsize\begin{tabularx}{\linewidth}{lXll}
\toprule
Comparison & Release date (2024) & Pair & Strict \\
\midrule
Q1 GDP $>$ GDI & 30 May, second & 1.3/1.5 & Does not pass \\
              & 27 June, third & 1.4/1.3 & Passes \\
Q2 GDP $>$ real PCE & 25 July, advance & 2.8/2.3 & Passes \\
              & 29 August, second & 3.0/2.9 & Passes \\
              & 26 September, third & 3.0/2.8 & Passes \\
\bottomrule
\end{tabularx}
\end{table}

Q1's GDP--GDI direction changes; the May release is an explicit counterworld.
The May and June BEA news releases supply the GDP/GDI paragraphs and revision
tables.\footnote{Source: U.S. Bureau of Economic Analysis, Q1 second estimate,
\url{https://www.bea.gov/index.php/news/2024/gross-domestic-product-first-quarter-2024-second-estimate-and-corporate-profits};
Q1 third estimate,
\url{https://www.bea.gov/news/2024/gross-domestic-product-third-estimate-corporate-profits-revised-estimate-and-gdp-industry}.
Accessed 20 October 2026.}
The advance release has no GDI estimate and is explicitly excluded from this
paired-complete task, not filled with zero. Q2 passes all three admitted modes,
although independent ranges cannot certify this: GDP's minimum 2.8 does not
exceed real PCE's maximum 2.9. That pair mixes July and August and is not an
admitted mode. Values come from the \emph{change} columns of SCB Tables~4
and~7, not adjacent GDP-contribution columns or the PCE price index.\footnote{BEA,
\emph{Survey of Current Business}, 5 September 2024, Table~4,
\url{https://apps.bea.gov/scb/issues/2024/09-september/0924-gdp-economy.htm};
4 October 2024, Table~7,
\url{https://apps.bea.gov/scb/issues/2024/10-october/1024-gdp-economy.htm}.
These articles reflect the August and September releases, respectively.
Accessed 20 October 2026.}
The repeated August pair agrees across both tables. September incorporates an
annual update, retaining reference year 2017; each vintage's own reported
quarter-on-quarter rates remain together. The two quarters are evaluated
separately, not assumed independent groups.

\paragraph{Restricted procedure.}
For fixed entity/window/pool, $r\geq1$, exact finite values and margins
$\delta_j$, the supported query is
$\exists c\in C,s_j\in S_j:\bigwedge_{j=1}^r c>s_j+\delta_j$.
There is at least one source group; each has a nonempty finite set of modes
and independently selects exactly one mode. Qualification and compatible
units are resolved beforehand. For mode $a$ of group $g$, construct
\[
 I_{ga,j}=[\max C_{ga},\;\min S_{ga,j}+\delta_j],\qquad
 F_j=\bigcap_g\bigcup_a I_{ga,j}.
\]
Absent roles give structural unbounded endpoints; discard reversed intervals.
A finite point $t\in F_j$ selects a covering original mode from every group:
all centres are at most $t$, and all comparator values plus margin are at
least $t$. This is a coherent non-positive world. Conversely, any world
failing comparison $j$ supplies such a separator, including absent-role cases.
The same maximal centre can serve every comparator, so the query is certain
exactly when every $F_j$ is empty. Ordinary colored interval coverage uses
$O(N+rM\log M)$ comparison operations for $N$ values and $M$ modes, apart
from exact-rational bit costs and output/checking costs. Cross-group
constraints, additional compatibility joins, changing windows and arbitrary
arithmetic are unsupported. This is a restricted specialization, not a new
geometric primitive or uncertainty semantics. Existing ProvSQL documentation
supports dependent numeric-aggregate comparisons.\footnote{ProvSQL,
\emph{Aggregation and Grouping},
\url{https://provsql.org/docs/user/aggregation.html}, accessed 20 October 2026.}

\paragraph{Execution and reproduction.}
Both cases agree with independent original-binding world enumeration and an
exact SAT encoding, and pass a separately recomputed output check. All five
modes are replayed. Fourteen test methods include 625 exhaustive and 400 seeded
controlled models plus application guards; these are software checks, not
additional real observations. The positive trace is ordinary execution
evidence, not a signed or succinct certificate. These one-group examples are
trivially enumerable: no practical speed advantage, native ProvSQL benchmark,
deployment or human benefit is established. Separate synthetic timing trials
are retained as prototype artifacts, not main empirical results.

Run \path{docs/jws_source_certainty_application.py}; inputs, bindings, outputs
and the exploratory protocol are in
\path{evaluation_jws/source_certainty_application_v1/}.
The \path{source_coherent_certainty_v1/} sibling retains synthetic checks.
Manual primary-page inspection supplies semantic/source review; executable
bindings check local transcription correspondence, not BEA's measurement
process. Hashes identify local artifacts, not publisher signatures. Attributed
numeric transcriptions follow BEA's public-domain statement, subject to stated
exceptions.\footnote{\url{https://www.bea.gov/help/faq/147}.}
These artifacts are part of the released revision archive.

\section{Objective evidence boundary}
The audit in \path{objective_study_v2/EVIDENCE_AUDIT.md} confirms that the private historical schemas contain no correctness-scored task judgments or durations. Public provenance and XAI participant datasets cannot estimate an OntoKG display effect because the tasks and allocation differ. Four instrument tests pass, but these are software tests, not participants. No new objective sessions have been conducted.
\section{Detailed source verification and acquisition history}
These paragraphs preserve the successive audit stages. A pending status describes its stage, not the latest consolidated evidence. In particular, the PPL completion overlay supersedes six earlier pending facts. The main paper reports current qualifications; historical inputs and reports remain unchanged.

\paragraph{Comparator authenticity is separate from arithmetic correctness.}
The PSX benchmark rows explicitly declare interpolation/forward-filling; MSX rows explicitly identify a damped selected-universe proxy rather than an official KLCI feed. Recomputing returns from those values does not authenticate a market benchmark. Classification results using these comparators are illustrative applications of the declared predicates, not verified market reactions. Source-linked arithmetic replay, disclosure-date alignment and authentic benchmark coverage must be reported independently. Official replacements must preserve the predeclared dates and horizon rather than substituting a convenient interval.

\paragraph{Bounded issuer-original authentication.}
A further purposive review checks 16 workbook facts against four issuer-authored documents. Four EPS cells match exactly in value and unit; four FFC monetary cells match exactly only after a factor of one million is applied to the workbook's declared thousand-PKR scale. Two ENGROH standalone PAT cells agree only at the issuer narrative's rounded million-PKR precision after scaling. Six OGDC/PPL annual sales, PAT and EPS cells cite half-year documents for December 2025 rather than the workbook year ended June 2025; these source-period mismatches are quarantined from authenticated-source claims. This is not a representative error-rate estimate. Reviewed transcriptions, exact cells, accounting scope, page references and source-copy hashes are retained separately. Automated reruns compare those reviewed transcriptions with workbook values; they do not independently read or authenticate the documents. Daily prices, FX, volumes and remaining market sources are not certified by this bounded review.

\paragraph{Versioned annual-source repair.}
The initial wrong-period audit is preserved. A subsequent repair authenticates six OGDC annual facts against issuer-original PSX document 259805, PDF page 3: FY2025 net sales and PAT are 401,177,969 and 169,903,614 thousand PKR, and EPS is 39.50 PKR; FY2024 comparative values are 463,697,861, 208,975,771 and 48.59. Four monetary cells require a factor of one million; two EPS cells match directly. Correct PPL annual-source candidates were located, but incomplete retrieval prevents visual financial-page authentication, so six PPL cells remain pending. The overlay distinguishes authenticated repairs from pending candidates and leaves archived inputs unchanged. Growth recomputation confirms common scaling cancels in the four reviewed ratios, but does not remove PPL's pending source status. FY2024 is checked against comparative columns, not separate original releases. OGDC's issuer letter dated 23 September 2025 is not treated as an authenticated first-publication timestamp or used to justify later windows.

\paragraph{Temporal quarantine rather than forced alignment.}
The separate temporal-resolution overlay excludes all 20 scaled R1 positives from report-event use: the workbooks explicitly label these anchors as recent-analysis windows, and their purported report dates repeat fiscal year-end. The TOWR mismatch is also quarantined. The 22 retained scaled positives are non-R1 classifications, not 22 authenticated earnings events. Bounded primary-source checks match Maybank's official FY2025 annual-report announcement of 1 April 2026 to its anchor (issuer announcement reference DCS-01042026-00004), and FFC's official financial-results document dated 29 January 2026 to its anchor (PSX document 269389). These are respectively an annual-report announcement date and an issuer document date, not authenticated first-earnings publication timestamps. Maxis remains declared-only: an AGM notice date is insufficient publication evidence. Neither fiscal period end nor an annual-report upload establishes the first earnings-release timestamp. No dates or missing price horizons are invented to recover excluded findings.

\paragraph{Bounded disclosure-dated cohort.}
A separate eligibility ledger retains all eight PSX/MSX companies, rather than selecting only successful cases. Exact observed-date Decimal replay gives three numerical positives and five negatives. Only FFC and Maybank have official dates matching their anchors; four candidates lack verified matching dates, and OGDC/PPL have wrong-period anchors without September 2025 price coverage. FFC's observed window runs from 30 January to 13 March 2026: company return is $-14.8750361\%$ versus supplied interpolated benchmark $-14.4853883\%$. Maybank's runs from 2 April to 14 May: company return is $-2.4432810\%$ versus proxy $0.3734215\%$. Each contains 30 closes and 29 shared intervals, beginning at the first post-disclosure close, not at a verified disclosure baseline. FFC's pre-anchor event closes are missing and Maybank has no pre-anchor company prices. Both R1 positives are therefore illustrative disclosure-dated replays, not authenticated initial reactions. No exact-date primary benchmark replacement pair was obtained in the bounded search. Eight controls pass, including rejection of missing intermediate company or benchmark dates rather than silent horizon truncation. These two cases are subsets of earlier cohorts, not additional independent positives.

\paragraph{Issuer-published company-price recovery.}
A subsequent source-acquisition pass recovered Maybank's public investor-relations CSV, containing 61 daily rows from 2 March to 29 May 2026. All 30 post-report company closes match the archived values exactly. Additional as-published closes are MYR11.360 on 31 March and MYR11.660 on 1 April; the 2 April and 14 May closes are MYR11.460 and MYR11.180. Thus the earlier local absence of a baseline is partially resolved by a separately hashed capture, without altering historical graphs. Company-only endpoint returns to 14 May are approximately $-1.58451\%$ from 31 March and $-4.11664\%$ from 1 April, compared with $-2.44328\%$ from the first post-report close. These are issuer-IR republished values, not exchange-original authentication. The export parameter \texttt{unadjusted=0} is unexplained by the viewed source, so adjustment basis remains unresolved. No compatible primary KLCI observations were obtained; neither a market-relative classification nor an initial earnings reaction follows from this company-only recovery. Source hashes, CSV field/line locators and four integrity controls are retained; redistribution rights are not assumed.

\paragraph{Regulator-published PSX endpoint recovery.}
Normal public attachment downloads subsequently recovered four original SECP reports for 28--30 January and 13 March 2026. Visual review confirms eight unique observations: six KSE100 levels and FFC closes of PKR597.10 on 29 January and PKR501.02 on 13 March. Repeated index values agree across adjacent reports. The FFC endpoints give a descriptive price change of approximately $-16.09111\%$, not a cash-distribution-adjusted return; the March row is explicitly marked XD. FFC is absent from the selected spot-price tables in the recovered 28 and 30 January reports, so neither its required pre-disclosure baseline nor its first-post close is filled by those documents. An opening price is not substituted for a previous close. KSE100's authenticated levels remain total-return observations, incompatible with the company price-only comparison; KSE100PR is still required. Thus original-source coverage improves, while the market-relative finding remains unvalidated. The earlier indexed-only ledger is retained as acquisition history; a separate source-hashed recovery ledger records the admitted observations and five integrity/refusal controls.

\paragraph{PPL completion.}
The complete nine-page issuer results letter was subsequently obtained from \url{https://www.psx.com.pk/psx/files-attachment/?file=259574.pdf}. Rendered page 3 confirms unconsolidated FY2025 revenue 242,516,363 and PAT 92,027,450 thousand PKR, and EPS 33.82; FY2024 comparatives are 288,797,413, 114,309,077 and 42.01. Four monetary cells need a factor of one million; two EPS cells match. Ten validation checks pass. This supersedes the earlier PPL pending status only through the separate completion overlay. The issuer letter date of 19 September 2025 remains distinct from first-publication time.

\end{document}